# Cooling in strongly correlated optical lattices: prospects and challenges

D. C. McKay[1] and B. DeMarco[1]

**Abstract**

Optical lattices have emerged as ideal simulators for Hubbard models of strongly correlated materials, such as the high-temperature superconducting cuprates. In optical lattice experiments, microscopic parameters such as the interaction strength between particles are well known and easily tunable. Unfortunately, this benefit of using optical lattices to study Hubbard models comes with one clear disadvantage: the energy scales in atomic systems are typically nanoKelvin compared with Kelvin in solids, with a correspondingly miniscule temperature scale required to observe exotic phases such as *d*-wave superconductivity. The ultra-low temperatures necessary to reach the regime in which optical lattice simulation can have an impact—the domain in which our theoretical understanding fails—have been a barrier to progress in this field. To move forward, a concerted effort is required to develop new techniques for cooling and, by extension, techniques to measure even lower temperatures. This article will be devoted to discussing the concepts of cooling and thermometry, fundamental sources of heat in optical lattice experiments, and a review of proposed and implemented thermometry and cooling techniques.

## 1. Introduction

Following work by Paul Benioff in 1980 on quantum mechanical models of computers as Turing machines [1], Richard Feynman delivered a talk during a Physics of Computation workshop in 1981 held at MIT that is credited for introducing the concept of *quantum simulation* [2]. He speculated that a universal quantum computer could efficiently simulate models of many-particle quantum systems that are beyond the reach of any classical computer, a conjecture that was later proven by Seth Lloyd in 1996 [3].

The problem that inspired Feynman is the exponential scaling of resources required to simulate a quantum system as the number of particles increases. For example, completely simulating the quantum state of 300 interacting spin-1/2 particles would require $2^{300} \approx 10^{90}$ bits of classical memory—a number larger than the estimated number of protons in the universe. While studying many-particle quantum systems using numerical simulation without a full-scale quantum computer may therefore seem hopeless at first blush, the situation is not quite so desperate. Efficient Quantum Monte Carlo (QMC) methods have been developed for simulating the ground state for models of a wide range of quantum materials, such as superfluid helium [4]. Much, though, is still out of reach. Consequences of the Pauli Exclusion Principle (i.e., the so-called "fermion sign problem") impede exactly simulating even the static properties of large collections of fermionic particles, such as the electrons in solids [5], especially when strong interactions are present. Exactly calculating the dynamics of more than a few tens of strongly interacting quantum particles is beyond the capabilities of today's most powerful supercomputers, and advances in computing consistent with Moore's Law enable the addition of just a few particles per decade [6].

---

[1] Department of Physics, University of Illinois at Urbana-Champaign.

These limitations have frustrated our efforts to understand a wide range of quantum systems, because we cannot resort to numerical simulation when traditional theory approaches fail to provide a complete picture (or to check approximations). Unfortunately, despite some proof-of-principle quantum simulation demonstrations using few-qubit quantum computers [7,8], large-scale quantum simulation as Feynman envisioned is likely to remain a challenge for some time.

Happily, developments in our ability to cool atomic gases to ultra-low temperature have potentially opened the door to circumventing the requirement of a full-scale quantum computer in certain cases. The principle is to use ultracold atom gases to simulate ideal models of other systems—by tuning physical parameters with high precision, many models of interest can be exactly realized [9]. Numerous suggestions for using ultra-cold gases as model systems have emerged over the last decade, from analogues of quantum chromodynamics [10] to paradigms for solids [11]. The modus operandi for quantum simulation in these experiments is to probe phase transitions between different quantum states of matter as experimental parameters are varied, thereby mapping out the phase diagrams of the corresponding model.

In this paper, we will focus on one problem in particular: using ultra-cold atoms trapped in an optical lattice to simulate variants of the Hubbard model. At the moment, much attention is focused in lattice experiments on cooling to low enough temperature to realize magnetically ordered states that are known to exist in the Hubbard model, and then reaching even lower temperature to discover if proposed superfluid states—the analogue of superconducting states in the cuprates—emerge. Achieving low enough temperature to probe such regimes of unknown physics has become considered a benchmark for the success of lattice simulation.

So far, achieving the temperature scale for magnetic ordering has proven to be out of reach. This article is dedicated to discussing the challenges that experiments have faced in cooling to lower temperature and the prospects for overcoming them. In the rest of this introduction, we review the basic physics of optical lattice experiments, how atoms trapped in lattices realize the Hubbard model, essential features of Hubbard models, the important differences between solids and lattice experiments, and state-of-the-art experimental tools. In Section 2, we discuss the concepts of cooling and thermometry important to lattice experiments and our current understanding of fundamental limits to cooling—i.e., is it physically possible to reach the low temperature regime of magnetic ordering? In Sections 3 and 4 we review the state-of-the-art in lattice cooling and thermometry and proposals for new techniques.

## 1.1. Simulating the Hubbard model

In optical lattice experiments, ultra-cold atoms are trapped in a crystal of light. Over the last decade, using optical lattices to study physics models, such as the Hubbard model, relevant to strongly correlated materials has generated tremendous excitement and a convergence of atomic and condensed matter physics [9]. The premise is to use optical lattices as an analogue for a solid material, with the atoms playing the role of the electrons (or superconducting electron pairs), and the light acting as the ionic crystal.

Variants of the Hubbard model have been used as paradigms for electronic properties of solids. In particular, the two-dimensional Fermi-Hubbard (FH) model has been proposed as a model for the high-temperature superconducting cuprates (see, for example, Ref. [12]). A proposed, schematic phase diagram of these materials is shown in Figure 1 [13]. Much is known about the basic features of the cuprates [14], such as the *d*-wave nature of the superconducting order parameter. However, while a great deal of the underlying physics has been revealed, we do not have a complete, microscopic picture for how high-temperature superconductivity emerges. Even phenomena at relatively high temperature, such as thermopower at room temperature [15-23] and transport in the "pseudogap" regime [24], remain poorly understood.

The Fermi-Hubbard model in its simplest form involves only two ingredients: particles tunneling between adjacent lattice sites with energy $t$, and particles in opposite states of spin on the same site interacting with energy $U$. When the interactions between particles are strong ($t/U < 1$), "strongly correlated" phases of matter can emerge that cannot be understood even qualitatively using any single particle theory (such as mean field theory). While the phase diagram of the FH model for repulsive interactions at "half-filling", or for a density corresponding to one particle per site, is well known, the nature of the FH model at lower fillings has been the subject of intense debate. For example, we are uncertain whether *d*-wave superconductivity exists in the FH model, as in the cuprates.

In 1998, it was pointed out in a theory paper that atoms trapped in an optical lattice realize the Hubbard model [25]. The advantage of using atoms to study these models is that the microscopic physics, such as the interactions between atoms, is very well understood and easily controllable. A grand challenge for the field that has developed is to use optical lattices to determine the conditions necessary for *d*-wave SF in the Hubbard model. The idea is to start with the atomic realization—two spin states of a fermionic atom trapped in an optical lattice— of the simplest FH model, cool to low temperature, and search for *d*-wave superfluidity (the analog of SC for neutral atoms). If the simplest FH model is insufficient to generate *d*-wave SF, then we can add in long-range interactions, disorder, and other features, and determine the impact on the phase diagram. Ultimately, the hope is to use optical lattices to measure the FH phase diagram.

Experimental and theoretical work on optical lattice simulation has not been focused solely on the FH model. An in-depth review of proposals can be found in Ref. [9]; here, we mention a few areas that lattices are primed to impact. Bosonic atoms trapped in a lattice realize the Bose-Hubbard (BH) model [26]. In the simplest, spinless BH model, particles tunnel between sites and interact if they are on the same site, just as in the FH model. The primary difference with the FH model are that the particles obey Bose statistics, and therefore particles in the same spin state can interact. While the ground state phase diagram of the BH model is well understood (see Refs. [27-29], for example), dynamics are not, and lattice experiments are beginning to have an impact on that front [30-36]. Adding disorder to bosonic particles in an optical lattice is a method for studying the disordered Bose-Hubbard (DBH) model [9,37,38], which has been used as a paradigm for granular superconductors and superfluids in porous media. In the DBH model, the characteristic physical parameters, such as the tunneling energy, vary from site-to-site. Experiments are starting to influence our understanding of the DBH model [39], about which there remain some disputes. Finally, ultra-cold atoms in a lattice can be used to study a variety of interacting spin models that involve magnetic interactions between spins pinned to a lattice (see, for example, Refs. [40-44]). Many of these models, particularly those involving frustration, remain unsolved.

## 1.2. Optical lattices

In optical lattice experiments, neutral atoms are first cooled to nanokelvin temperature as a gas while confined in a parabolic potential (characterized by a harmonic oscillator frequency $\omega$). Bosonic atoms (e.g., $^{87}$Rb, $^{7}$Li, $^{23}$Na), fermionic atoms (e.g., $^{6}$Li, $^{40}$K), or a combination can be used. An optical lattice potential is superimposed on the gas by slowly turning on a combination of laser beams. The simplest lattice potential is realized by using pairs of counter-propagating laser beams with identical polarization and wavelength $\lambda$ (Figure 2). Each pair creates an intensity standing wave and corresponding periodic potential $V_{lat}\sin^2(2\pi x/\lambda)$ through the AC Stark effect, with the potential depth $V_{lat}$ proportional to the local light intensity. To create a cubic lattice, each direction has an orthogonal polarization so that the lattice potentials along each direction add independently, creating an overall potential $V_{lat}\left[\sin^2(2\pi x/\lambda)+\sin^2(2\pi y/\lambda)+\sin^2(2\pi z/\lambda)\right]$. Square lattices can be made by making one pair high intensity in order to confine the atoms into a series of "pancakes"; similarly, two strong pairs can trap the atoms in a series of tubes to create an ensemble of one-dimensional lattices (Figure 2). A commonly used notation is to specify the lattice potential depth as $s = V_{lat}/E_R$, where $E_R = h^2/2m\lambda^2$ is the "recoil energy" ($m$ is the atomic mass and $h$ is Plank's constant).

While a wide variety of crystal geometries are possible, the nature of the AC Stark effect limits what optical lattice potentials are achievable [45]. For a monochromatic laser field and an alkali atom, the potential from the AC Stark effect is:

$$U_{dip}(\vec{x}) = \frac{\pi c^2 \Gamma}{2\omega_0^3}\left[\left(\frac{2}{\Delta_{3/2}} + \frac{1}{\Delta_{1/2}}\right)I(\vec{x}) + g_F m_F \sum_{q=-1,0,1} q\left(\frac{1}{\Delta_{3/2}} - \frac{1}{\Delta_{1/2}}\right)I_q(\vec{x})\right]$$

[46] where $c$ is the speed of light, $\Gamma$ is the decay rate of the electronic excited state, $\omega_0$ is the angular frequency of the atomic transition, $I_q$ is the laser intensity with polarization $q$ ($I = \sum_{q=-1,0,1} I_q$), and $\Delta_{3/2}$ ($\Delta_{1/2}$) is the detuning $\omega_L - \omega_0$ relative to the $S \to P_{3/2}$ ($S \to P_{1/2}$) transition ($\omega_L$ is the angular frequency of the laser); for a more in-depth discussion see Ref. [47]. This potential is in the rotating frame defined by the laser frequency and is only correct for laser detunings large compared to the atomic hyperfine splittings; if multiple laser frequencies are employed, the overall potential is not necessarily the sum of the potential generated by each laser field. In general, the multi-level, multi-field problem must be solved to determine the overall potential. In Sec. 2.3 and the appendix we discuss how this formula must be modified for very large detunings ($|\omega_0 - \omega_L| \gtrsim \omega_0$). Various lattice geometries are possible using different laser beam configurations. To date, atoms have been trapped in in the strongly correlated limit in cubic [26], square [48-50], one-dimensional [51], hexagonal [52,53], and triangular [54] lattices. Atoms have also confined in spin-dependent lattices (for which the laser detuning must be comparable to the atomic fine structure) that involve polarization gradients in the strongly correlated regime [47,52,55,56].

In a lattice, the atoms develop a band structure, just like electrons in a solid. The atomic wavefunctions can be described as superpositions of Wannier states, where a single Wannier state, $w_i(\vec{x})$, describes an atom localized on the $i^{th}$ potential well of the lattice. For certain geometries (see Ref. [57], for example), the Wannier states are straightforward to calculate. The tunneling energy for atoms to hop between adjacent sites is $t = \int d^3x\, w_i^*(\vec{x})\left[-\frac{\hbar^2}{2m}\nabla^2 + V(\vec{x})\right]w_j(\vec{x})$, where $i$ and $j$ label adjacent sites, and $V(\vec{x})$ is the lattice potential (somewhat unfortunately, $t$ and $J$ have been used interchangeably in the literature; we reserve $J$ for the super-exchange energy). In the discussion that follows and for the rest of this paper, we ignore excited bands and consider only the lowest energy band (a good approximation for $s \gtrsim 4$). For single particles in a uniform, one-dimensional lattice, the resulting Hamiltonian is $H = -t\sum_{\langle ij\rangle} \hat{a}_i^\dagger \hat{a}_j$, with a spectrum $E(q) = 2t[1 - \cos(\pi q/q_B)]$, where $\langle ij\rangle$ indicates a sum over adjacent sites, the operator $\hat{a}_i^\dagger$ ($\hat{a}_j$) creates (destroys) a particle in state $w_i(x)$ on site $i$, $q$ is the atomic quasimomentum, and $q_B = \hbar\pi/d$ is the Brilloun-zone momentum ($d$ is the distance between sites). For a detailed discussion of quasimomentum, see Ref. [58]. The term "bandwidth" is often used to refer to the range of energies $4t$ in the band.

The parabolic confining potential present in experiments modifies the spectrum and quantum states [58,59]. As in a finite solid system, the spectrum becomes discrete; furthermore, the bandgap disappears. The low energy states are harmonic oscillator states modulated by the lattice potential, while at higher energy the states become localized to the edge of the lattice. In Sec. 2.3, we will employ the effective mass approximation, which is valid for both trapped and uniform systems for low energy states and $s \gg 1$. In this approximation, the effect of the lattice is only to renormalize the mass according to $m^* = \hbar^2/2d^2t$, where $m^*$ is referred to as the effective mass. For trapped gases, the harmonic oscillator frequency is modified according to $\omega^* = \omega\sqrt{m/m^*}$.

The interaction between atoms on the same site drastically modifies this single particle picture. Atoms at these low temperatures interact primarily through an *s*-wave collision. The interaction energy between two atoms on the same site is approximately $U = \frac{4\pi\hbar^2 a_s}{m}\int d^3x |w(\vec{x})|^4$, where $a_s$ is the atomic scattering length (we note that interactions may affect the Wannier states, leading to occupation-based corrections to the Hubbard parameters [60-62]). Bosonic atoms in the same spin state can interact; in order for fermionic atoms to collide they must have different states of spin. By "spin" we mean the hyperfine state of the atom, which is specified by the quantum numbers $F$ and $m_F$, and consists of the combined electronic and nuclear total angular momentum. In typical lattice experiments, the number of atoms in each spin state is fixed, in contrast to electronic systems. We will ignore inelastic, spin-changing collisions between atoms. This physics can play an important (or dominant) role in lattice experiments [63], although not in the context we envision here. For studying the FH model, for example, typically two hyperfine states are selected to proxy for spin up and down. The relative populations are determined experimentally by adjusting the relative number in each state.

These two ingredients—tunneling and on-site interactions—exactly realize the Hubbard model. If the atoms are (spin-polarized) bosons, then the BH model is realized:

$$H = -t\sum_{\langle ij \rangle} \hat{a}_i^\dagger \hat{a}_j + \frac{U}{2}\sum_i \hat{n}_i(\hat{n}_i - 1) + \sum_i \epsilon_i \hat{n}_i$$

where $n_i = \hat{a}_i^\dagger \hat{a}_j$ is the number of atoms on site $i$ and $\epsilon_i = m\omega^2 r_i^2/2$ is the potential energy (from the parabolic confinement) for site $i$ located at radius $r_i$. Or, if the atoms are fermions, the FH model:

$$H = -t \sum_{\langle ij \rangle, \sigma} \hat{a}^{\dagger}_{i,\sigma} \hat{a}_{j,\sigma} + U \sum_i \hat{n}_{i,\uparrow} \hat{n}_{i,\downarrow} + \sum_{i,\sigma} \epsilon_i \hat{n}_{i,\sigma}$$

where $\sigma = \uparrow, \downarrow$. A thermodynamic chemical potential $\mu$ is usually introduced to fix the total particle number. We will discuss how corrections (arising from interactions) to these single-band models can be important in lattice experiments.

All of the parameters in the Hubbard model that control the behavior of the system can be relatively easily tuned in an optical lattice experiment. The ratio $t/U$ can be tuned by changing the lattice laser intensity, which changes the lattice potential depth; increasing the potential depth decreases $t$ and increases $U$. The interaction energy can also be adjusted independently using a Feshbach resonance. The density can be controlled by tuning the confinement and/or the number of atoms.

As $t/U$ and the density are varied, different quantum phases emerge at zero temperature in a uniform system. For spin-polarized bosons, the BH model gives rise to superfluid and Mott-insulator phases depending on the density (Figure 3). For low interaction energy, the particles delocalize into a superfluid (SF) state, while above a critical interaction strength a transition to a localized Mott insulator (MI) phase occurs for integer fillings (i.e., number of particles per site). Because of the confining potential, which is often treated using the local density approximation (LDA) and an effective chemical potential $\tilde{\mu} = \mu - m\omega^2 r^2/2$, the phases from the uniform system appear inhomogeneously. For example, for bosons, at high lattice depths (and low temperature) the lattice is filled with nested Mott-insulator and superfluid phases (see inset to Figure 3). This structure was originally detected using microwave spectroscopy [64] and spin-changing collisions [65], and recently directly imaged using high-resolution microscopy [36,66,67].

In the LDA, the phases present in the lattice are understood by sampling a vertical line on the homogeneous phase diagram (with $\mu/U$ and $U/t$ as axes). The characteristic density $\tilde{\rho} = N\left(m\omega^2 d^2 / 2t\right)^{j/2}$ ($j$ is the dimensionality) is a useful quantity—along with $U/t$—as an alternative for characterizing quantum phases in a trapped system [68-70]. The characteristic density can be used to convert an LDA phase diagram into a universal one for which the state of the system is characterized by a single point. By specifying single values of $U/t$ and $\tilde{\rho}$, one can uniquely determine the phases present and their spatial arrangement (Figure 4). Furthermore, this approach enables the inclusion of trap-induced modifications to critical phenomena that go beyond the local-density approximation [71].

The characteristic density is especially helpful for understanding the phases of the inhomogeneous FH model (Figure 4). For low interaction strength and low density, a delocalized metallic Fermi liquid (FL) state exists. Instead of a FL, a band insulator forms if the filling is two particles (one spin up, one spin down) per site. As with bosons, above a critical interaction strength (in this case, determined by the bandwidth), the transition to a Mott insulator of fermions occurs. For a recent review of studying these FH phases using optical lattices, see Ref. [72].

For systems with more than one spin-component, there is still a spin degree of freedom in the localized MI phase that can lead to new phases not shown explicitly in Figure 4. Unless $t = 0$, the atoms are not completely localized and small, but finite, hopping events—referred to as virtual tunneling—give rise to another energy scale

$$J = 4t^2/U,$$

which is the super-exchange energy. Super-exchange is known to play an important role in magnetic phenomena in solids and was probed experimentally using atoms confined in an array of double wells [42]. For the two-component Hubbard model with one atom per site, equal tunneling and interaction energies for all components, and $t/U \ll 1$, the Hamiltonian reduces to

$$H = \pm J \sum_{\langle ij \rangle} \vec{S}_i \cdot \vec{S}_j,$$

which is known as the Heisenberg spin Hamiltonian ($\vec{S}$ is the spin operator) [40,41,73], where the sign is positive (negative) for fermions (bosons). For fermions, this Hamiltonian gives rise to an anti-ferromagnetic (AFM) ground state (i.e., alternating spin up-spin down ordering) at zero temperature because the particles can slightly lower their energy in this configuration via virtual tunneling. Reaching the regime in which AFM order begins to emerge, below the Neel temperature $T_N = J/k_B$, is a primary goal for fermion lattice experiments and will be a necessary first step on the way to the regime of *d*-wave SF. A number of other magnetic Hamiltonians can also be simulated if the tunneling and/or interactions can be made state and /or directionally dependent [40,41]. The energy scales of these magnetic phases are very low, and the techniques required to detect and achieve these phases will be a large part of the discussion in subsequent sections.

We think it is important to mention a few differences between solids and optical lattices. A primary difference between lattice experiments and solids is precisely the variation of density across the trap. In solids, the electron density is roughly constant and controlled by doping. In lattice experiments, the density is highest in the center of the lattice. The interaction between atoms is naturally short ranged, in contrast to the long-range Coulomb interaction between electrons. The equivalent of phonons and other lattice distortions are absent in an optical lattice. Effects in solids arising from inner-shell or multiple outer shell electrons are also missing in optical lattice experiments we discuss here, and so overlapping bands do not play a role.

## 1.3. State-of-the-art Experimental Tools

Experimental progress on using optical lattices to study variants of the Hubbard model has been rapid. We mention only a few relevant highlights here. A bosonic Mott insulator has been realized in cubic [26], square [48-50], one-dimensional, hexagonal [52], and triangular [54] lattices. The BH phase diagram has been measured in a square lattice, and a small but definite deviation from the LDA was detected [71]. The Mott insulator phase for fermions has been achieved in a cubic lattice [74,75]. Notably, temperatures low enough for magnetic ordering or potential *d*-wave SF have not been reached, although super-exchange oscillations have been measured between adjacent sites [42]. Mixtures of atomic species in the strongly correlated limit have been prepared in a lattice, including a Fermi-Bose mixture of $^{40}$K and $^{87}$Rb [76-78], and a Bose-Bose mixture of $^{41}$K and $^{87}$Rb [79].

A number of unique techniques have been demonstrated for creating lattice potentials. Lattices have been created by holographic projection using phase masks [80,81], imaging micro-lens arrays [82], and magnetically [83]. The polarization of the lattice beams can be used to create spin dependent lattices [47,52,55,56] and also a lattice of double wells [84]. Spontaneously emitted light from atoms trapped in an optical cavity can then give rise to a lattice [85]. Another approach is to implement superlattices using more than one wavelength of light [43]. If the wavelengths are incommensurate, the potential is quasi-disordered and the atoms are described by the Aubry-André Hamiltonian [86]. Dipole traps can be added to compensate the harmonic potential of the lattice beams [61]. Light sent through a disordered phase mask can be used to apply a speckle potential to the lattice, which is a method to implement the disordered Bose-Hubbard Hamiltonian [38]. Dynamical lattices have been implemented to explore the properties of models with time-varying parameters. These include rotating lattices [87,88], lattices with rotating wells [89], and position-modulated lattices [90].

A panoply of tools for measuring properties of atoms trapped in a lattice has been developed theoretically and experimentally. The quasimomentum distribution can be measured, with some limits at high quasimomentum [58,91,92]. Analyzing the noise correlations in a set of time-of-flight images reveals the second order momentum correlations, which can probe the SF-MI transition and detect Bose and Fermi bunching/anti-bunching effects [93,94] The equivalent of many measurements on solids have been demonstrated and are possible, such as the excitation spectrum [74], transport [16,30-32,95-97], and compressibility [74,75]. Site and atom-number resolved imaging has recently been demonstrated [36,67]. The excitation spectrum can be measured using Bragg and Raman spectroscopy [98,99]. If AFM ordering or *d*-wave superfluidity is present, there are a number of realistic theoretical proposals for detecting it (see Secs. 4.3 and 4.4).

## 2. Cooling and thermometry

We have made a strong case that optical lattice experiments are poised to address fundamental questions about paradigms from condensed matter physics and phenomena involving strong correlations. The experimental tools have been in place for several years to realize a wide variety of models, to control the analogue of material parameters, and to characterize and detect different quantum phases. So what, then, is preventing experiments from probing, for example, the regime of AFM and *d*-wave superfluidity in the Hubbard model?

The stumbling block is temperature—both our inability to measure it and to reach far enough into the ultra-cold frontier. Much of the recent history of AMO physics has been dominated by the goal of cooling atoms to ever lower temperatures in order to explore quantum many-body physics and to improve precision measurements. Reducing absolute temperature is the important function of cooling for the latter application. For example, cooling Cesium atom gases to microKelvin temperatures enables long interrogation times and commensurate high precision for the fountain atomic clock, which is our current time standard.

In the context of the present discussion, cooling is a process used to lower the entropy per particle $S/N$. Quantum phase transitions are controlled by entropy because it measures the number of accessible quantum states. Entropy is generically a complicated and often unknown function of the number of particles, temperature, interactions, and the confining potential. Therefore, lowering absolute temperature is not necessarily sufficient to reduce $S/N$. An example of a method—adiabatic expansion—that has been used to reduce $T$ to as low as 450 pK for a trapped gas [100] without affecting $S/N$ is shown in Figure 5. So, while it is convenient to refer to how cooling reduces temperature, and we will use this language, for the remainder of this paper by "cooling" we mean a technique that decreases $S/N$.

A triumph of work during the 1980s and 1990s on cooling the motional degrees of freedom of atomic gases was the achievement of quantum degeneracy of both bosons and fermions [101,102]. While a wide range of cooling methods were proposed and realized during this quest, the workhorse of most experiments remains a relatively simple combination of laser and evaporative cooling (Figure 5). Laser cooling is effective at cooling atoms gases from room temperature where $S/N \sim 40 k_B$ to the microKelvin regime, for which $S/N \sim 10 k_B$, typically. A remarkably powerful process that can provide over 1 kW of cooling for the human body, evaporative cooling is used to further reduce $S/N$ to below 3.6 $k_B$, which, in a parabolic trap, corresponds to the critical temperature $T_c$ for condensation for ideal (i.e., non-interacting) bosons and approximately 1/2 the Fermi temperature $T_F$ for ideal fermions. Time-of-flight (TOF) thermometry (Figure 6) played an essential role as a tool for optimizing cooling in this quest for quantum degeneracy. Temperature has also been a quantity of physical interest, such as in early measurements of the damping of collective modes [103] and in more recent studies of dissipation in optical lattices [30].

In this section, we lay the groundwork for the comprehensive discussion in Sections 3 and 4 of the theoretical and experimental state-of-the-art for thermometry and cooling in optical lattice experiments. We introduce the fundamentals of these topics both by reviewing past work and making forays into other fields. In particular, we pay attention to a key concept that has perilously been ignored in the literature: cooling into low entropy states can only be successful if the cooling power exceeds the heating rate in the regime of interest. We also explain the complications for cooling and thermometry created by strong correlations. Finally, we address in detail some issues we believe to be universal to these experiments, especially the fundamental limitations to cooling generated by the very light used to create the lattice (which has also not been thoroughly investigated).

## 2.1. Measuring temperature

In optical lattice experiments, accurate thermometry will be necessary to reach the ultra-low $S/N$ required to realize novel phases. Furthermore, a precise determination of temperature will be a key ingredient in using optical lattice experiments to experimentally determine the Hubbard model phase diagram. In the weakly interacting regime, temperature is typically inferred for harmonically trapped gases by imaging the integrated density profile after turning off the confining potential and allowing the gas to expand (i.e., TOF imaging). For sufficiently long expansion time, the column density profile is equivalent to the integrated momentum profile, which can be fit to certain hypergeometric functions to determine temperature, as shown for a Fermi gas in Figure 6.

Interactions are the primary complication for TOF thermometry. The expansion can significantly deviate from ballistic even for moderate interaction strength (i.e., the interaction energy per atom is comparable to the kinetic energy). Indeed, in most BEC experiments the gas expands hydrodynamically [104]. Interactions also change how observables translate into temperature by, e.g., modifying the equation of state [105] or by distorting the effective potential experienced by the atoms [106]. Fortunately, theory for nearly all experiments not involving an optical lattice are under enough control such that measurements of the density profile after TOF can still be connected directly to temperature. For example, in BEC-BCS crossover experiments, the profile can be fit to the non-interacting result and temperature can be inferred using the known equation of state [105]. Even without theory, temperature can often still be determined in this regime from the "tail" of the distribution, which corresponds to high kinetic energy states with low occupation. Unfortunately, this procedure fails at very low temperature, when the signal-to-noise ratio in the tail of the distribution is too poor to obtain a reliable fit.

Determining the temperature in optical lattices in the strongly correlated regime faces a fundamental, rather than technical, problem. Ultimately, we are interested in probing the regime for which we have no verified or well controlled theory. Therefore, we will, in general, lack a method for connecting observables—such as any part of the density profile after TOF—to temperature. We may also be interested in the temperature of other degrees of freedom, such as spin, for which we have no proven general technique (although methods for spin thermometry in certain limited regimes are emerging [107]). The challenge is thus two-fold: to develop thermometry methods that do not rely on unverified theoretical results and that can be experimentally validated. Experimental validation requires achieving consistency between two or more techniques.

This situation is analogous to problems associated with thermometry in cryogenic experiments with solids, as summarized by Pobell in Ref. [108]. In that arena, *secondary thermometers*, such as $RuO_2$ resistance thermometers, are the workhorse of experiments. These thermometers are calibrated against *primary thermometers*, such as measurements of thermal noise in a resistor or the angular anisotropy of gamma rays emitted from nuclear isomers that are products of $\beta$-decay. In both cases, a separate material may be used as the thermometer and contacted to the sample. A primary thermometer is one which we can connect the measured quantity to temperature from first principles. For ultra-cold atom gas experiments, TOF thermometry (for weak interactions) is an example of a primary thermometer.

Thermometry in cryogenic experiments faces numerous technical challenges, similar to the issue of finite imaging signal-to-noise ratio in TOF thermometry. For example, any particular thermometer has a limited regime of operation (e.g., the sensitivity may be poor below or above some temperature), and so an array of thermometers must be used to span the temperature range of interest. The thermometer may "self heat", thereby heating the sample and defeating attempts to reach low temperature. Also, a thermometer may lose thermal contact with the sample. Or, more subtly, the electrons in the sample may be out of thermal equilibrium with the phonons, which transfer heat between the sample and thermometer. This problem and the related issue of the sample losing thermal contact with the cryostat are common at low temperatures. Consequently, there may be ambiguity regarding the root cause of a measured quantity saturating or nonlinear behavior as one attempts to cool the sample (see Refs. [109] and [110], for example). At best, consistency between several thermometers may be monitored as a check on thermal contact.

We feel that there are several important lessons from thermometry in cryogenic applications. Foremost, the challenge of primary thermometry in lattice experiments originates in employing intrinsic properties of the gas for measuring temperature. If we are to venture into regimes for which we have no complete theory, then creating a thermometer understood from first principles will necessarily be problematic. Experiments on solids overcome this issue by using a different material with well understood properties as a thermometer. For ultra-cold lattice experiments, we must therefore develop a similar extrinsic thermometer, or create multiple primary thermometers based on different theoretical approximations, and then check for consistency at low temperatures. The progress that has been made on both of these fronts will be discussed in Sec. 4.

We should also be mindful of the analogue of technical issues from the cryogenic realm. For extrinsic thermometers, we must ensure thermal contact and equilibration on relevant timescales. Any extrinsic thermometers should have a low heat capacity and not introduce heat to the gas of interest. Developing secondary thermometers can be useful, as long as they can be calibrated against a primary method. All thermometers should have high sensitivity to changes in temperature, and so we will likely need different methods in different regimes of temperature.

In Section 4 we will assess the state-of-the art of thermometry in light of these guiding principles. For the remainder of this section, we will focus on exploring the ultimate limits to temperature in optical lattice experiments.

## 2.2. Temperature limits in optical lattices

In current optical lattice experiments, the lattice potential is slowly superimposed on the atom gas after it is first cooled to as low temperature as possible in a purely parabolic potential. The goal is to make the lattice turn-on as adiabatic as possible so that $S/N$ is preserved. The lowest attainable $S/N$ in a harmonic trap is therefore a lower limit to what will be achievable in a lattice. The published state-of-the art in cooling trapped gases reaches lower bounds of 0.05 $T/T_F$ for non-interacting gases [111]. For strongly interacting gases, similar *effective* temperatures (derived from fitting TOF distributions to a non-interacting profile) have been measured [105,112], and a careful study demonstrated cooling to $S/N \sim 0.6\ k_B$ [113]. We caution that measuring temperature in the strongly interacting regime requires modeling [113,114], and measuring temperatures below 0.1 $T/T_F$ tends to be dominated by systematic errors related to imaging [114,115]. For the weakly interacting (in a harmonic trap, before transfer into a lattice) regime relevant to lattice experiments, 0.13 $T/T_F$ [116] and 0.3 $T/T_c$ [30] are the lowest reported temperatures for fermions and bosons respectively, corresponding to $S/N \sim 1 k_B$ and $S/N \sim 0.1 k_B$ for an ideal gas. For fermionic atoms, this is too high to realize the anti-ferromagnetic phase that exists below $S/N \sim k_B 0.5 ln 2 \sim 0.35 k_B$ [70,116-119]. To study low-energy physics of the Hubbard model and access *d*-wave paired states will require cooling to yet lower $S/N$.

We now arrive at the crux of the matter: in order to realize low-entropy phases such as an AFM in optical lattice experiments, much lower temperatures are required. For non-interacting fermions, $S/N = -k_B \left[ 24 (T/T_F)^3 Li_4(-\mathfrak{z}) + \ln \mathfrak{z} \right]$ (where the fugacity $\mathfrak{z}$ is determined by $Li_3(-\mathfrak{z}) = -(T/T_F)^{-3}/6$ and $Li_n$ is the polylogarithmic function of order $n$), and therefore reaching $S/N \sim 0.5 k_B ln 2$ will require $T/T_F < 0.04$. It may be possible to reach lower than $S/N \sim 0.5 k_B ln 2$ in a lattice by starting at even smaller entropy in the parabolic potential. But, methods for cooling to far lower entropy in parabolic potentials may not be sufficient, because non-adiabaticity during the lattice turn-on and heating from the lattice may be too severe. In Section 2.3 we argue that this is likely the scenario for reaching below the super-exchange temperature scale for bosons in a lattice because of unavoidable heating from the lattice light during the lattice turn-on. As we will also discuss, reaching low entropy for fermions may be frustrated by longer than expected adiabatic timescales.

The challenge is then to develop methods to cool atoms *in a lattice* to $S/N \sim 0.35 k_B$. Understanding the limitations of cooling—i.e., what limits the ultimate achievable temperature—is of key importance to evaluating the many different proposals for lattice cooling that we review in Section 3. The limitation for any cooling scheme can be understood as a competition between heating and cooling. Because the temperature during cooling evolves according to $\dot{T} = \dot{T}_{heat} - Q$, all cooling schemes are limited by the same condition: the lowest temperature possible is achieved when the cooling power $Q$ equals the heating rate $\dot{T}_{heat}$. Heating is unavoidable in ultra-cold atom experiments and arises both from instrinsic and technical sources. How exactly this condition plays out depends on the details of the heating present and the cooling method employed. Furthermore, atom loss, either unintended or purposeful as in evaporative cooling, will generally lead to a limit on $S/N$ occurring at $Q > \dot{T}_{heat}$ since $\dot{S} = (C\dot{T} - \mu\dot{N})/T$ ($\mu$ is the chemical potential and $C$ is the heat capacity).

To better understand the interplay of heating and cooling power, we use forced evaporative cooling as an example. Evaporative cooling has been exhaustively studied for both classical and quantum gases confined in parabolic potentials. Typically, it is modeled as a process that truncates the trapping potential at an energy $\eta k_B T$; atoms with higher energies are ejected from the trap. For large evaporation parameter $\eta$, atoms with above the average energy and $S/N$ are lost, thereby resulting in cooling.

For evaporative cooling at constant and high $\eta$, $Q \propto N$, so the cooling power is reduced as atoms are lost and the temperature drops. An example of an evaporative cooling "trajectory" is shown in Figure 7, calculated using the kinetic model from Ref. [120] for a classical gas (i.e., one obeying Maxwell-Boltzmann statistics) with an atom loss time constant $\tau$ (from, e.g. collisions with residual gas atoms). Cooling fails below 250 nK under the chosen scenario, coincident with $Q < \dot{T}_{heat}$ (note that conditions are generally superior to this in realistic experiments). The limiting $S/N \approx 7k_b$ occurs for $Q > \dot{T}_{heat}$ at the point in the trajectory when $C\dot{T} = \mu\dot{N}$. Although evaporative cooling has no in principle temperature or $S/N$ limit, finite heating and loss rates always lead to a practical limit, just as in this example. Because not all heating and loss sources can be eliminated—particularly those arising from the lattice light—the design and evaluation of any proposed cooling method must include a comparison of the cooling power with a realistic heating rate.

For fermionic atoms there are notably two other inescapable limitations to any cooling method—hole heating and Pauli blocking. Hole heating arises from atom loss, which may result from collisions with residual gas atoms (as indicated by time constant $\tau$ in the example in Figure 8) or from density-dependent losses (e.g., three-body recombination, dipolar loss, or spin-exchange). Entropy is produced as atoms are lost, with collisions continually repopulating low-energy states, leading to the promotion of atoms to high-energy states and rethermalization to higher temperature. The heating rate associated with loss at temperatures much lower than the Fermi temperature is severe and given by $\dot{T}_{heat} = (4/5\pi^2)(T_F^2/\tau T)$ [121]. For a gas cooled to $0.01 T_F$, the temperature doubling time is 0.2% of $\tau$, so cooling power on the order of at least $100 T_F/\tau$ is required to reach temperatures this low. For a system with $T_F \approx 200$ nK and $\tau = 100 s$, this requires $Q > 200$ nK/s.

This intrinsic heating process is especially problematic when combined with Pauli blocking,

which limits the cooling rate at low temperature [122,123]. Pauli blocking affects all dynamical processes in Fermi gases, including rethermalizing collisions necessary to cooling. Collisions can be understood as a phenomenon that re-arranges the particles among the energy levels of the system (Figure 9). The average collision rate per atom is proportional to the density of final states for each colliding partner. Cooling requires one of the fermionic colliding partners to emerge after the collision in a state at lower energy. At temperatures far below $T_F$, most low energy states are occupied and are unavailable as final states because of the Pauli exclusion principle. Therefore, the collision rate is reduced, tending to zero as $(T/T_F)^2$ [124]. Because the collision rate is a limiting timescale for cooling ($Q$ for evaporative cooling is proportional to the collision rate, for example), $Q$ is always reduced to zero as the gas is cooled.

In summary, there is an important conclusion to be drawn from our discussion of evaporative cooling. In order to assess the efficacy of any cooling method, we require realistic models for cooling and heating power, since the lowest achievable temperature is determined by the competition between positive and negative heat flow. In fact, naively one would assume that evaporative cooling could reach arbitrarily low entropies unless realistic heating and loss rates were included in a theoretical model. Unfortunately, we generally lack sufficient models because we do not have a quantum Boltzmann treatment—necessary to capture dominant effects at low entropy such as Pauli blocking and hole heating—that includes strong correlations. Therefore, it is likely that *progress on the theory of dynamics and thermalization in strongly correlated systems will have a strong impact on guiding experiments to cool into new regimes.*

We will examine proposed cooling methods in Sec. 3 in light of this discussion of cooling power and quantum statistical effects. First, though, we wish to examine what is likely to be the largest stumbling block to reaching lower entropy—an unavoidable source of heating resulting from the interaction of the atoms with the light used to create the lattice. As we will discuss, this rate is temperature independent, and so a higher cooling power (compared to the thermal energy) will be required to combat it at low temperatures. This heating has not been dealt with extensively in the literature, and so, in the next section, we discuss its properties and likely impact on reaching low $S/N$.

## 2.3. Light-induced heating

All optical lattice experiments are afflicted with an intrinsic heating rate that arises from the interaction of the atom with the light that creates the optical lattice potential. Gordon and Ashkin [125] first approached this problem by calculating the momentum diffusion rate $D_p = \frac{1}{2}\frac{d}{dt}(\langle \vec{p}\cdot\vec{p}\rangle - \langle \vec{p}\rangle\cdot\langle \vec{p}\rangle)$ for a particle with momentum $\vec{p}$ in a standing wave formed from counter-propagating laser beams. Somewhat counter-intuitively, they found $D_p = \hbar^2 k^2 \Gamma^3 \tilde{I}_{max}/16\Delta^2$ for large detuning, with $\tilde{I}_{max} = I_{max}/I_{sat}$, where $I_{max}$ is the intensity at an anti-node and $I_{sat}$ is the saturation intensity. The diffusion rate does not depend on the location of the atom, i.e., whether it is trapped at a node (as in a blue-detuned lattice) or an anti-node (as in a red-detuned lattice). We prefer to consider $D_p$ as deriving from two contributions: atomic recoil arising from spontaneous emission, and the force from zero-point fluctuations of the atomic electric dipole interacting with gradients in the magnitude of the standing-wave electric field (Figure 10). The zero-point contribution, which has often been ignored, can be understood in the following way. Fluctuations in the vacuum electric field polarize the atomic electric dipole; this induced dipole interacts with electric field gradients to produce a force on the atom. While the average dipole moment and force vanish at the standing wave nodes, the average of the dipole moment squared is finite, and hence momentum diffusion occurs even in dark regions of the light field. The recoil contribution vanishes at the nodes (where the light intensity vanishes), whereas the zero-point component is zero at the anti-nodes (where the standing-wave electric field has no gradient). Gordon and Ashkin's unexpected discovery was that the zero-point component at a node is exactly equal to the recoil contribution at an anti-node, and, moreover, that the total diffusion rate is constant at all points in the standing wave.

The diffusion rate can straightforwardly be converted into an energy dissipation rate $\langle \dot{E} \rangle = D_p/m = E_R V_{lat} \Gamma / \hbar |\Delta|$, where $V_{lat}$ is the lattice potential depth. We use $m$ instead of $m^*$ in this equation because momentum diffusion occurs on the timescale of the electronic excited state lifetime [125], which is many orders of magnitude faster than any relevant lattice timescale. For a single standing wave and a harmonically trapped particle, the associated heating rate is $\dot{T} = \langle \dot{E} \rangle / 3k_B$ (assuming that the energy thermalizes equally into all three dimensions). In the discussion that follows, we use this formula to estimate a heating rate, primarily for simplicity. It is correct in the effective mass limit, in which the standing wave does not affect the heat capacity; the effective mass approximation is valid for low kinetic energy even given the rapid momentum diffusion rate, since thermalization times are typically long compared with the tunneling time.

More recent papers [126,127] improve on Gordon and Ashkin's calculation using a master-equation approach for the quantized atomic motion in a lattice. Ref. [126] considered a single atom, whereas Ref. [127] studied a many-body calculation for bosons. In a limit obeyed by all current experiments, these master-equation approaches find the same result for the total increase in mean energy as Gordon and Ashkin—namely, that the mean energy increase is independent of the sign of the lattice detuning. Both Refs. [126] and [127] question whether the total energy increase is the relevant metric. For high lattice potential depth, heating from the lattice light is mostly related to transitions to higher bands. Since energetic considerations and 1D simulations [127] imply the atoms in higher bands do not decay to the ground band and thermalize on experimental timescales, Refs. [126] and [127] focus on scattering events that do not result in an inter-band transition. When only ground-band heating events are considered (i.e., atoms promoted to higher bands are artificially removed from the system), the heating rate is higher for red-detuned versus blue-detuned lattices (at the same magnitude of detuning). In our experience, however, decay rates from higher bands are rapid (on the order of several milliseconds) in cubic lattices, so this may be a significant source of heating, and therefore more investigation is required.

Given that the key question is how lattice light heating affects entropy, *it is imperative to*

*develop a many-body quantum mechanical description of light scattering in a densely populated optical lattice* in order to accurately assess and to overcome the impact of heating. The many-body calculation for single component bosons in a lattice in Ref. [127] is an important first step along these lines. The main result of that work is to highlight that scattering events localize atoms to specific lattice sites. For the superfluid state, which is delocalized, this causes fast decay of the off-diagonal long range order in the correlation function. The Mott insulator state, which is already localized, is more robust against scattering. Since this effect depends on the overall *rate* of scattering, it is worse in red than blue detuned lattices.

While Ref. [127] is a preliminary stride towards a full understanding of the heating effects from the lattice light, it is worth noting several areas which still need to be addressed. In a densely populated lattice with a high optical depth, multiple scattering events may affect the recoil contribution to the heating rate in much the same way Bragg scattering strongly influences fluorescence imaging [128]. Also, spontaneous Raman scattering into other hyperfine ground states was ignored—while suppressed for large detuning [129], such scattering may alter the heating rate, especially for magnetic degrees of freedom. Finally, effects important for detunings comparable to the resonant frequency [46] were ignored. Furthermore, interactions and quantum statistics may change the heating rate by modifying the density of states [130-133]. *Most importantly*, Ref. [127] must be extended to address fermionic and multi-component gases. Ref. [127] only considered bosonic superfluid and Mott insulator states, yet the main challenge moving forward is to cool into fermionic anti-ferromagnetic phases. Addressing heating in this regime is vital.

Although calculations such as in Ref. [127] are needed to fully assess the heating rate, the simple heating model from Ref. [125] has been demonstrated to be *roughly* consistent with experiments [134]. We will therefore use this straightforward model for relative comparisons across a wide range of parameters in order to address the impact of light-induced heating on experiments to probe Hubbard phenomena using optical lattices. Two questions arise immediately—first, how do we vary experimental parameters such as the atomic mass $m$, lattice spacing $d$, and scattering length $a_s$ to minimize the impact of the heating while keeping the physics of interest unchanged? And, second, what is the relevant figure of merit?

With regards to the first issue, we claim that, for both fermions and bosons, a fair comparison at different $m$, $d$, and $a_s$ can be made by keeping the characteristic density $\tilde{\rho}$, $U/t$, and the ratio of scattering length to lattice spacing $a_s/d$ fixed. By specifying single values of $U/t$ and $\tilde{\rho}$, one can uniquely determine the phases present and their spatial arrangement (Figure 4). The ratio $a_s/d$ determines the size of corrections to the single-band Hubbard model description of an optical lattice [60,61,135-137]. We note that the ratio of $U$ to the bandgap energy $E_{bg}$ can also be used as an equivalent measure [137]— $U/E_{bg} \propto s^{1/4} a_s/d$ (which scales very weakly with $s$). As $a_s/d$ grows, the tunneling and interaction energies become increasingly density dependent; by fixing $a_s/d$, we can constrain these corrections to be small and constant.

For the rest of this section, we therefore concentrate on exploring the impact of heating for a fixed point on the phase diagram determined by a specific $\tilde{\rho}$ and $U/t$, keeping $a_s/d$ constant. We choose $U/t=18$ for fermions and $U/t=45$ for bosons (set by the lattice potential depth), ensuring in each case that the center of the trap consists of a unit filled MI phase. We also choose $a_s/d=0.01$ (maintained in these calculations by assuming a Feshbach resonance is available and adjusting $a_s$—potentially difficult for some atoms such as $^{87}$Rb [138]), which leads to negligible corrections to $t$, for example, for unit filling and the range of parameters considered here. For fermions, this combination of parameters ensures that we sample the single-band Heisenberg regime, and that we are close to the maximal value of the super-exchange energy $J$ (according to dynamical mean field theory [70,117]). We note that there are indications that $J$ may be maximal for significantly lower $U/t$, and therefore improvements in the lattice-induced heating may be possible for fermions [135]. We restrict our attention to cubic lattices formed from three pairs of laser fields with mutually orthogonal wave-vectors and polarizations. In the appendix, we discuss modifications to the energy dissipation rate for lattices with imperfect contrast and for lattices formed using laser fields that intersect at an angle.

We choose the laser wavelength (and therefore the lattice constant) and atomic species as parameters to vary. Even though changing $m$ and $d$ will likely affect cooling power, we feel that it sensible to consider the heating rate independently since it is not apparent which cooling method will ultimately reach the low entropies of interest. We consider two bosonic species ($^{87}$Rb and $^{133}$Cs) and two fermionic species ($^6$Li and $^{40}$K) that span a wide range of atomic masses. We also consider wavelengths from 400–1550 nm, corresponding to 200–775 nm lattice spacings.

Now we are poised to address the second question: what is a relevant figure of merit? Unfortunately, there is no single figure of merit that can capture the impact of heating in lattice experiments—there are multiple energy and timescales involved, and preparation of quantum phases may proceed using fundamentally different methods. In spite of this, some general conclusions can still be reached about the impact of heating, methods for minimizing heating, and the likelihood that heating will prevent experiments from reaching low enough entropy to realize, e.g., anti-ferromagnetic phases. In Table 1 we collect several formulae (valid in the limit $s \ll 1$) and definitions useful for this discussion. For this formulae and the plots in this section, we apply corrections to $\dot{E}$ (as derived in Ref. [125]) and the lattice potential depth relevant for large detunings (see the appendix and equations 10 and 11 in Ref. [46] for a treatment of a two-level system). This includes a correction for the counter-rotating term in both the dipole potential and heating rate, and a modification for the photon density of states in the heating rate.

| Heating rate | $\dot{T} = \dfrac{h^3}{2k_B cm^2} \dfrac{s\Gamma}{\lambda^3} \left(\dfrac{\lambda_0}{\lambda}\right)^3 \left|\dfrac{1}{\lambda/\lambda_0 - \lambda_0/\lambda}\right|$ |
|---|---|
| **Tunneling energy** | $t = 8\pi^{3/2}\hbar^2 \dfrac{s^{3/4}e^{-2\sqrt{s}}}{m\lambda^2}$ |
| **Interaction energy** | $U = 8\sqrt{2}\pi^{5/2}\hbar^2 \dfrac{a_s s^{3/4}}{m\lambda^3}$ |
| **Super-exchange energy** | $J = 4t^2/U = 16\sqrt{2\pi}\hbar^2 \dfrac{s^{3/4}e^{-4\sqrt{s}}}{m\lambda a_s}$ |
| **Ratio of Hubbard energies** | $U/t = \pi\sqrt{2}\,\dfrac{a_s e^{2\sqrt{s}}}{\lambda}$ |
| **Characteristic timescale** | $\tau_t = h/t,\ \tau_J = h/J$ |
| **Figure of merit** | $F_{Ut} = \dfrac{\dot{T}}{k_B}\Big/\left(\dfrac{U}{\tau_t}\right),\ F_{tt} = \dfrac{\dot{T}}{k_B}\Big/\left(\dfrac{t}{\tau_t}\right),\ F_{JJ} = \dfrac{\dot{T}}{k_B}\Big/\left(\dfrac{J}{\tau_J}\right) \propto \dfrac{\lambda_0^3}{\lambda^2}\dfrac{1}{|\lambda/\lambda_0 - \lambda_0/\lambda|}$ |

Table 1. Formulae useful for assessing the impact of light-induced heating.

It is tempting to use the absolute heating rate $\dot{T}$, shown in Figure 11, as a figure of merit. The heating rate is minimized for high mass species and longer wavelengths, primarily because this reduces the recoil energy. Naively, then, the optimal experiment would work with either $^{133}$Cs or $^{40}$K in with a lattice with the largest possible lattice spacing. However, characteristic energies such as $t$, $U$, and $J$ shrink for longer wavelengths and higher masses. Therefore, to understand the impact of the heating rate, we cannot examine solely its absolute value. Rather, one should compare the heating rate to the ratio of a characteristic energy to time.

Formulating a characteristic time for lattice experiments is problematic, and strongly depends on the approach used to prepare low-entropy states. In general, we assume that experiments will be dominated by the longest relevant timescale: either $\tau_t = h/t$ if tunneling or kinetic energy plays a dominant role, or $\tau_J = h/J$ if magnetism arising from super-exchange is of interest. These timescales will determine equilibration times for excitations and adiabatic turn-on times of the lattice. Therefore, they are appropriate to describe schemes that attempt to reach low entropy by initially cooling to low temperature and then adiabatically turning on the lattice, cooling in the lattice, or some version of filtering. We emphasize that understanding the process by which certain excitations equilibrate, such as spin waves, is a current topic of research, and it is not clear if this simple analysis is appropriate.

We can now formulate a figure of merit $F_{\epsilon\tau}$ based on a characteristic energy $\epsilon$ and time $\tau$. We choose three combinations. For experiments probing a MI, we choose $F_{Ut}$, since the relevant energy scale is $U$ and tunneling will determine adiabatic transformation and equilibration times. For measurements of SF or FL properties, we choose $F_{tt}$, since $t$ sets the scale of $T_c$ and $T_F$ as well as the characteristic timescale. Finally, for experiments designed to probe AFM, we suggest $F_{JJ}$, because $J$ is the relevant energy scale and spin-excitations are likely to relax on a timescale related to $\tau_J$.

Perhaps most important is the qualitative dependence of these figures of merit on the parameters that can be controlled experimentally. All are independent of the atomic mass, since the heating rate is proportional to $1/m^2$, and each energy and time scale is proportional to $1/m$. Therefore, although the heating rate is highest for light species such as Li, this should not strongly impact the ability to reach low entropy. Also, all figures of merit are proportional to $\dfrac{\lambda_0^3}{\lambda^2}\dfrac{1}{|\lambda/\lambda_0 - \lambda_0/\lambda|}$, which implies that the impact of heating is minimized at long lattice spacings and for short wavelength electronic resonances. This is evident in the plots of $F_{tt}$, $F_{Ut}$, and $F_{JJ}$ shown in Figure 12.

Qualitatively, we see that $F_{tt}$ and $F_{Ut}$ can be smaller than unity across a wide range of laser wavelengths for all species. This should not be surprising, considering the realization of superfluidity and the Mott-insulator in optical lattices for both fermions and bosons. The situation is a somewhat different for $F_{JJ}$. For fermions, $F_{JJ}$ can be well below one for laser wavelengths far enough from the electronic resonance. One way to interpret this is that if the gas was instantaneously cooled to zero temperature, the time to heat above the superexchange energy scale is much slower than the dynamical timescale associated with, e.g., spin-waves. *We conclude, therefore, that as long as sufficient cooling power is available, there is no barrier in principle to accessing the AFM state for fermions.* This does not imply that any potential *d*-wave SF state may be within reach. In the cuprates, *d*-wave superconductivity occurs at temperatures 3–4 times lower than the undoped AFM state [13]. Not only, then, will the filling have to be controlled in optical lattice experiments, but even lower entropy will be required.

The situation for bosons is somewhat less encouraging with respect to studying magnetic phenomena induced by superexchange— $F_{JJ}$ is of order one or greater across the full range of accessible wavelengths for bosonic species. The reason $F_{JJ}$ is higher for bosons compared with fermions is that stronger lattices are required to reach the MI regime (again, keeping $a_s/d$ constant). This consequence of quantum statistics will likely foil efforts to observe super-exchange induced magnetic phenomena for bosons.

In conclusion, adiabatic transfer into an AFM state may be possible for fermions if lower $S/N$ can be reached for parabolically confined atom gases. The margin in the figure-of-merit $F_{JJ}$ that measures the degree to which adiabaticity is limited by light-induced heating is approximately a factor of 10 for the lattice laser wavelengths employed in current experiments ($1064\ nm > \lambda > 700 nm$). Additional sources of heating, i.e., from technical noise, are therefore a concern, as are indications that adiabatic timescales may be longer than expected. During attempted adiabatic transfer, evidence is mounting that mass transport may limit adiabaticity [96,139]. Additionally, thermalization timescales may be long depending on how excitations decay [140]. Finally, reaching the low entropy states of interest will require thermalization between the spin and motional degrees of freedom, which has not yet been demonstrated. There are some promising approaches that may obviate these limitations, such as magnetic lattices [83]—a method for avoiding limitations from light-induced heating. Ultimately, we feel that in-lattice cooling methods will be necessary, particularly to reach any potential *d*-wave SF state; we discuss such cooling proposals in the next section.

## 3. State-of-the-art Lattice Cooling

It is clear that new entropy reduction techniques are needed if the quest for low entropy phases is to be successful. Simply attempting to circumvent the limits discussed in the previous section, by, for example imposing antiferromagnetic ordering in isolated wells and slowly increasing the coupling may not work [141]. Although it may be possible to explore certain phases, such as AFM, as the highest energy states of certain Hamiltonians [142], our focus will be on cooling to the ground state. In the following we will review a number of proposed entropy reduction methods. Although no significant changes to current experiments are required, all but one (spin-gradient demagnetization) remain to be implemented. The proposals fall roughly into two groups. Included in the first group are filtration schemes, which take advantage of entropy residing in certain modes of the system which can then be "filtered" out. The second group is based on immersing the system in a reservoir that carries away the entropy.

We note that cooling power has not been calculated for any of these proposals. Unfortunately, the notion that the success of any cooling method can only be assessed via a comparison of cooling to heating power has been largely absent from the literature on lattice cooling techniques. As we pointed out in Sec. 2.2 using evaporative cooling as an example, such an evaluation is necessary. We comment on the implications of this oversight at the end of this section and remark on what work will be necessary to move forward.

### 3.1. Filter cooling

There are several different types of filtering schemes proposed. The first method, which we will refer to as *spatial filtering*, utilizes the harmonic trap to create high entropy regions which can be removed from the system. With appropriate tuning of the confinement, a gapped phase occurs in the center of the trap (i.e., a band insulator for fermions) and the majority of entropy will reside at the edge (Figure 13a). This high entropy region can be filtered from the system by adding a potential barrier [143] or displacing [144] or significantly weakening [145] the potential at the edge. Simulations suggest a reduction in entropy per particle to $S/N \approx 0.1$ k$_B$ [143] and $S/N \approx 0.001$ k$_B$ [145]. An advantage of spatial filtering schemes is that they require minimal changes to be implemented in current experiments. For the scheme suggested in Ref. [143] the main limitation is the ability to create effective barrier potentials, which has recently been demonstrated, albeit not in application to cooling [36]. The filtering method proposed in Ref. [145] relies on adiabaticity between a very weakly trapped outer region and a strongly confined inner region. It is likely that the timescales for maintaining this adiabaticity are unrealistic and further calculation of these is required.

The next category, *band filtering*, involves transferring entropy to higher bands; atoms in these higher energy states are then removed. In fermionic systems, atoms occupy higher bands when the first band is full because of the Pauli exclusion principle. Since there are no available states in the lowest band, it has close to zero entropy, and so the majority of the entropy resides in higher bands [146-148]. This is even more effective if traps can be tailored with flat potential profiles (and hard walls) [147] since there are no true bands in harmonic systems [58,59]. A possible method for removing the higher bands is to use Raman transitions to free particle states [147]. Simulations of non-interacting fermions suggest that temperatures as low as $T/T_F \approx 0.001$ can be attained if the proper potentials can be created [147]. To further evaluate this method, we need to understand the role of interactions, particularly during the removal of the atoms in the higher energy bands.

The final set of proposals, *number filtering*, are based on transferring entropy contained in number fluctuations into another internal state [149-151] (Figure 13b). The other state can then be removed from the lattice, or the two states decoupled via a Feshbach resonance or by shifting the lattice [152]. In deep lattices, entropy is mainly carried by number fluctuations if only a single component is present. These schemes are based on the fact that, although near the edge of the trap the fluctuations are between empty and occupied sites, in the centre the fluctuations are between states with finite occupancy. By engineering schemes to transfer exactly one atom per site to a different internal state, for example, these fluctuations can be eliminated. This scheme only works if the occupancy is non-zero, so it is best suited for bosonic systems in tight traps where there is high occupation at the centre. Calculations of this process show that the entropy per particle can be reduced by a factor of 2 after one filtering operation. However, by relaxing and then re-squeezing the trap (which carries entropy from the edges to the centre), and repeating the filtering step, the entropy can be rapidly reduced in as little as four cycles. A complication is that once number fluctuations are removed, the system is not in equilibrium, so the trap must be relaxed to prevent entropy regeneration. A limitation of this method is that it only works for specific ratios of interactions, and a more realistic analysis of the timescales for carrying the entropy from the edge to the centre is required to understand the efficiency of applying this scheme cyclically [139].

Beyond the three main filtering proposals, there are a few others that we will briefly summarize. The first is *algorithmic number filtering*, which creates a final state similar to number filtering as described above, except the system is filtered in a step-wise fashion. For example, if access to both single site imaging and single site addressability is available, defects can be repaired by individually moving atoms to empty sites [153,154]. Another procedure is to split the system in two, separate the two parts, and bring them slowly back together again. As the edges of the two gases start to overlap, atoms are removed, for example by using a Feshbach resonance [149,150]. This has the effect of sharpening the number distribution at the edge of each gas. Another proposal is a *dynamical filtering,* in which removing the harmonic confinement causes low entropy regions consisting of doubly occupied sites (in the Fermi-Hubbard model) to collect together [155]. A means to recapture this region has not been proposed, however. The final proposal is to *spectrally filter* by applying a potential gradient to spatially map energy to density [156]—a process known as a spectral transform. This procedure allows the system to be filtered by removing atoms from the sites corresponding to higher energies. Subsequently, the spectral transform is reapplied, taking the system back to the original lattice eigenstates. However, the effect of harmonic confinement, the timing of the transform, and how the system rethermalizes after evaporating and reapplying the transform requires more investigation.

## 3.2. Immersion cooling

The next set of *immersion cooling* proposals are based on immersing the system to be cooled (the "sample") in a "reservoir" system that can carry away entropy (Figure 14a). The most straightforward proposal is to adiabatically transform the potentials of the system in such a way that the entropy of the reservoir increases, and therefore the entropy of the sample decreases. An experimental realization of this scheme was investigated in a harmonic trap [157], where it was shown that by compressing the sample independently of the reservoir that the sample, which was bosonic, could be reversibly condensed. There are similar proposals to explore these effects where only the sample experiences the lattice, using different states of the same species [47], or two different species [158,159]. The most detailed study [158] considered a fermionic system in a lattice as the sample, and a harmonically trapped BEC as the reservoir. By increasing the confinement of the sample heat flows into the BEC reservoir, and calculations show a decrease to $S/N \approx 0.02$ $k_B$. There are also a number of theoretical and experimental studies where both species experience a lattice potential [76-79,160-162]. Given species-dependent interaction and tunneling parameters, there can be an increase in entropy in the one at the expense of the other, which may be exploited for cooling purposes. The primary remaining issue is to understand the thermalization rate in these systems. For example, in Ref. [158] the simulated entropy decrease requires full adiabaticity between the BEC reservoir and the sample. However, if thermalization is not efficient, then the intrinsic heating and loss processes will dominate.

An interesting variation on this concept is to use two degrees of freedom of the same gas to play the role of the reservoir and the system. For example, if two spin components of the same species are co-trapped, the spin degree of freedom can be used as a reservoir for the motional/position degrees of freedom. Such a cooling scheme is common in condensed matter systems and is known as adiabatic demagnetization cooling: the sample is polarized in a high magnetic field, and, as the field is lowered, the spins disorder and absorb entropy. In a cold atom system, rapid spin-changing but non-spin-conserving collisions—such as those behind dipolar relaxation—are required for this method. So far, this cooling technique has only been demonstrated for the non-alkali Cr at relatively high temperature and for a gas that was purely harmonically trapped; Cr has a high magnetic moment and therefore a high dipolar relaxation rate [163]. A related technique—spin-gradient demagnetization cooling—has been demonstrated using atoms trapped in a lattice [164]. Here, a two-spin Bose gas is prepared in a magnetic field gradient, which segregates the components to opposite sides of the trap. The width of the inter-species mixing region can be used to measure the spin temperature, as discussed in Sec. 4.5. As the gradient is lowered, the two species mix, and ideally entropy from, for example, particle-hole excitations is transferred into the spin mixing entropy. There are indications of cooling via comparison to theory [164], but direct evidence is lacking since there is no way to independently verify that spin and particle-hole excitations are coupled. This type of cooling will be limited by magnetic correlations when the temperature is on the order of the superexchange energy [164,165], and therefore may be of limited usefulness for accessing the AFM state.

A way to circumvent this limit is discussed in Ref. [166], which proposes to cool into the AFM regime using a high-spin species. In this scheme, a mixture of two spin states selected from a larger manifold is cooled in a parabolic trap to low temperature in the presence of a large quadratic Zeeman effect. Subsequently, a lattice is applied and spin-changing collisions are enabled by decreasing the magnetic field. Then, an inhomogeneous quadratic Zeeman effect is applied, and entropy is segregated into a spin liquid at the edge of the gas. While this scheme will work into the regime of strong magnetic correlations, limiting timescales may reduce the effective cooling power below the threshold required for accessing the AFM regime.

Another route to immersion cooling is to drive the sample to an excited state, such as an excited band, and let it decay by releasing an excitation into the reservoir [167-169] (Figure 14b). Cooling can occur if the system is able to decay to a state lower in energy than the initial state. Simulations show that for a non-interacting Bose system an effective temperature of $k_B T/4t \approx 0.002$ [169] can be attained in a time $50/t$. Practical issues, such as interactions between lattice atoms and interactions during the excitation and re-absorption of excitations may limit the effectiveness of this approach. A general concern for all immersion proposals is that we lack the experimental demonstration of a lattice system immersed in a BEC reservoir, although the problem is currently being studied [47,157].

## 3.3. Conclusion

As discussed in Section 2.2, the effectiveness of any cooling scheme must be evaluated by comparing cooling power to heating rates. In contrast to evaporative cooling, many of the schemes presented here proceed in a step-wise fashion, and so only a time-averaged cooling power can be defined. Unfortunately, cooling power (including quantum statistical effects, such as Pauli blocking) in the strongly correlated limit has not been calculated for any of the proposed cooling schemes. Ultimately, all these cooling schemes will be limited by the timescale for the coupling between the atomic motion and spin through super-exchange [158], since we wish to cool into a low-entropy phase in which the motional and spin degrees of freedom are in equilibrium. In principle, then, any of these schemes may work for fermions, as we discussed in Section 2.3. Without more information such as the cooling power and the scale of technical heating, however, we cannot assess the likelihood of success in practice. That said, any scheme that relies on state-dependent potentials will be fundamentally limited to a poor figure of merit since the light must be detuned on the order of the atomic fine structure splitting. Species-dependent potentials suffer a similar problem for at least one of the species, but certain atom combinations may limit the heating for the species of interest [159]. For example, in the scheme of Ref. [167] heating from the light potential on the reservoir species should not affect the temperature of the lattice gas being cooled.

To summarize, we have discussed several cooling schemes which work to transfer entropy out of the system of interest, by either filtering the entropy or immersing the system in a low entropy reservoir. Progress is being made towards the goal of in-lattice cooling, including a (possible) demonstration of immersion cooling using a spin reservoir (i.e., spin-gradient demagnetization cooling). Another important step is that high-resolution potential shaping has been realized [36], which is a critical component for spatial filtering schemes [143,145,147]. It is clear from our discussion that the most important questions that remain to be addressed are related to the dynamics in lattice systems. This includes thermalization times between the system and the reservoir [158,164], the time required for transforming from the phases useful for these cooling schemes into the strongly correlated phases of interest, and the time for equilibration between the spin and motion. Finally, it remains to be seen whether a cooling scheme can be devised that directly removes spin entropy.

## 4. State-of-the-art Lattice Thermometry

In this section, we review the state-of-the-art in lattice thermometry, including methods that have already been employed and proposals for new techniques. Even though most existing methods cannot be extended into the low-entropy regimes of interest, they may provide guidance in developing new techniques. In this section we will be careful to distinguish between methods that are primary thermometers—those for which the measured quantity can be connected to temperature via first principles—and secondary thermometers, which require calibration. We note that only two of the methods we will discuss qualify as primary thermometers: measuring fluctuations via in-situ imaging and using a second weakly interacting species as a thermometer.

All thermometry methods can roughly be separated into five basic strategies. The first strategy is to assume that the lattice is turned on adiabatically (i.e., without a change in entropy) and therefore use traditional methods to measure entropy before the lattice is turned on. The second approach is to use theoretical input to understand TOF imaging. The third method is to measure in-situ distributions of the system. Another strategy looks at the response of the system to external perturbations. The final tactic is to develop extrinsic, primary thermometers—independent systems (or degrees of freedom) in thermal contact with the system under study.

### 4.1. Isentropic assumption

The standard method of preparing gases in optical lattices is to transfer the gas from a harmonic trap by turning on the optical power slowly compared with all timescales. If technical noise plays no role, then the entropy per particle in the lattice is limited only by the light-induced heating accumulated during the lattice turn on. For most experiments to date, the resulting fractional increase in entropy per particle is small, and $S/N$ before the lattice turn on is used to estimate the temperature in the lattice [74,75,116,170]. Clearly, this method will be of limited use as experiments pursue ultra-low entropy phases for which the entropy accumulated during the lattice turn on is significant.

To use this technique for thermometry, theory must be employed to connect entropy to temperature. The relationship between temperature and entropy can be calculated from theory in the non-interacting limit [59,146,171-174], with mean field methods [70,117,119,175-178], analytic approximations in certain limits [118,179], and with QMC in certain regimes [180-182]. The main limitations of this method are the assumption of adiabaticity, which is violated by heating processes in the lattice [116,125,170], and the reliance on theory, which makes this method a secondary thermometer in the strongly correlated regime. Heating can be estimated experimentally by measuring the entropy before loading into the lattice and after turning the lattice on and then off again [74]. A detailed study of these heating processes for bosons concluded that the main heating contribution is light-induced from the lattice beams and that the ramping process itself is adiabatic [170]. However, other studies have shown that while local adiabaticity is achieved during the ramp, global adiabaticity requires timescales an order of magnitude longer than typically used [139]. While the results of this thermometry appear to agree well with observations [74,75,116,170], more research is needed into the adiabaticity of the ramping process. This thermometry has a large range, but the lower bound is set by entropies that can be measured in the harmonic trap—$S/N \approx k_B$ (fermions) [74,75,116] and $S/N \approx 0.1\ k_B$ (bosons) [30]. This method will not be useful for evaluating in-lattice cooling schemes.

## 4.2. TOF imaging

There are a number of thermometry techniques that can be applied to atoms in the lattice. The first group of these is to analyze the images after releasing the atoms from the lattice (TOF imaging). This is a natural choice given that it is the main technique used for harmonically trapped systems and is not demanding of imaging resolution. Assuming that interactions play no role during the expansion and a long enough expansion [183], the TOF images probe the momentum distribution. In harmonically trapped, non-interacting gases, thermal energy is equally shared between position and momentum degrees of freedom (i.e., potential and kinetic energy), but this is not the case for lattice systems. Instead there is a limited amount of thermal energy that can be stored in momentum (i.e., kinetic energy), proportional to the tunneling $t$, which decreases exponentially with lattice depth. Once the band is nearly filled, the momentum distribution is not a sensitive probe of temperature [58]. For fermions, the limit corresponds to $T_F > t$ or $T > t$, and therefore TOF images are usually not used to probe the temperature of fermions in a lattice. For example, in three dimensions with sufficient atom number to achieve half-filling at the center of the trap [116], $T_F \sim 2.3t$ (in the non-tunneling limit [184]). Only for very weak traps and/or low atom number can TOF images be used for fermion thermometry.

Even when $T$ is less than $t$, TOF images can be difficult to interpret due to strong interactions. In the thermal regime or in the SF regime of the BH diagram, non-interacting fits to all or parts of the image can be used [50,58]. There have also been attempts to develop analytic, but approximate distributions [50,185], but these have not been evaluated in an experiment. The only unequivocal method is direct comparison of TOF images to QMC simulations [134], which also take into account finite expansion effects [183] (Figure 17b). Direct comparison to QMC simulations can be considered a primary thermometer for bosonic systems, but, in general, QMC simulations are not possible for interacting fermion systems. Despite being a primary thermometer, QMC comparison is limited because of the required computational power. It would be difficult to use this technique for evaluating the temperature of a running experiment, for example, to understand if the system is being properly cooled. By the time QMC results are complete, the experimental situation may have changed (i.e., due to ambient magnetic fields). The utility of this method therefore lies in its ability to calibrate secondary thermometers. Already in [134], QMC comparison was used to calibrate thermometry based on the isentropic assumption.

Secondary thermometry will undoubtedly play an important role cooling to low entropy. In this method, heuristic observables are defined in the TOF images which are then calibrated to temperature using either theory or a different thermometer (i.e., such as in Ref. [134]). One suitable observable is the visibility after TOF (Figure 17a) [186]. At zero temperature the visibility decreases as the lattice depth is increased and interactions fill the Brillouin zone—an effect that was used as evidence for the first experiments probing an optical lattice MI [26]. The visibility is temperature dependent above the critical temperature for bosons [180,181,187] and could be used for thermometry [185,186]. A "kink" (i.e., discontinuous derivative) in the visibility is one method for determining $T_c$, but the experimental error is large [134] (Figure 17c). A similar observable is to define a "condensate fraction" by identifying a bimodal distribution [187] that consists of a narrow peak on a broad distribution (Figure 17a). The fraction of atoms in the sharp feature is the heuristic condensate fraction. The condensate fraction defined in this way can be compared to theory [173,174,176-178,181,187] for thermometry below the critical temperature. The critical temperature can be identified by a vanishing condensate fraction or changes in the width of the narrow peak [50,170] (Figure 17c). Although locating the critical point does not perform thermometry, this achievement is important for calibrating and checking thermometry methods.

Finally, there are techniques that look at the statistical properties of a set of TOF images. For example, some of the noise in each image is a consequence of second-order momenta correlations. Correlations in this noise can be used to probe temperature in the MI phase for fermions and bosons [94], although the sensitivity has not been explored. In the AFM phase a peak should appear in the correlation at a special value of the momentum due to the formation of singlet pairs [188]. The weight of this peak could be a method for probing temperature in the regime of AFM ordering.

### 4.3. In-situ Imaging

In the previous section we discussed thermometry accomplished by analyzing TOF images. As mentioned, for deep lattices, temperature information in the momentum distribution is reduced since the energy that can be stored in momenta fluctuations scales as $t$, which shrinks exponentially as the lattice depth increases. Thermal energy for strong lattices is predominately stored in particles at the edge of the trapping potential, particle-hole excitations, and spin disorder (at low temperatures). These excitation modes appear weakly in TOF images as, for example, small signals in the noise correlations [82]. However, all these excitations are nearly diagonal in the site and spin-resolved atom number basis, which points to in-situ imaging as a fruitful method for thermometry. Although a few in-situ techniques do not require high-resolution imaging, it is necessary to exploit the full range of possible thermometry methods. This daunting technical challenge has been recently overcome by a number of groups [36,66,67,189]. Site resolved in-situ imaging has also been demonstrated for 5 μm lattice spacing [190] in 3D, 2 μm [82] and 600 nm [191] in 2D and 433 nm [189] in 1D for systems not in the strongly correlated regime.

In-situ imaging is well-suited for systems that are harmonically trapped. For probing techniques that lack position resolution, such as TOF imaging, the signal is averaged over all chemical potentials present in the trap. Consequently, measured properties may be averaged over several different phases, which can cause considerable issues of interpretation. However, for in-situ techniques with the ability to resolve position, the harmonic trap is an advantage. Because the chemical potential is small near the edge of the trap, a perturbative expansion for the density distribution [192] can be used to extract the temperature. A caveat is that the region for which this expansion is applicable grows smaller with higher interactions and lower temperatures, making finite signal-to-noise ratio a limitation. Also, incomplete global thermalization may lead to the local temperature at the edge of the gas differing from the local temperature in the center of the lattice [139,165]. An extension of this idea is to use higher order expansions [193], which are still numerically tractable and increase the region of applicability in the trap.

This approach can also be used to extract an equation of state, by first measuring pressure as the integral of density with respect to chemical potential (along the camera line of sight) [192]. By changing the harmonic confinement and temperature, the entropy can be determined experimentally from the derivative of pressure with temperature [192]. This method has been used to measure the equation of state for a strongly interacting Fermi gas, but not in a lattice [194,195].

The trap can also be utilized to determine the compressibility from the derivative of the density with respect to chemical potential (i.e., position) [66], which can be used to identify insulating phases. Furthermore, the ratio of compressibility to local number fluctuations (which can be directly measured) can be used to determine temperature via the fluctuation-dissipation theorem [193,196]; however, experimental errors using this method are large [66]. The fluctuation-dissipation theorem has also been used to measure the temperature of a degenerate Fermi gas in a harmonic trap with errors that are comparable to more traditional techniques [197,198]. Determining entropy and temperature using these two techniques (via the equation-of-state or fluctuation-dissipation theorem and measurements of the in-situ density), qualify as primary thermometry since the methods are entirely model independent. However, they both assume that the LDA is valid and that the system can be described using a grand canonical potential.

Another in-situ technique is to fit the distribution to an exact equation in the $t \to 0$ limit, which may be justified for deep lattices in which $U/t \ll 1$ [66,139]. In Ref. [66], the temperature was measured by fitting to in-situ distributions with approximately 20% error (see Figure 16). As cooling enables the realization of spin ordering, in-situ techniques will be able to identify these phases directly. Measuring the magnetic order parameter, such as the magnetization or staggered magnetization for antiferromagnetism, should be effective for spin thermometry. It is an outstanding technical challenge to demonstrate spin-sensitive imaging, however, there are a number of proposals (i.e., Ref. [199], and see also Sec. 4.4). Eventually, in-situ techniques will be limited when there is almost complete magnetic ordering and samples will be "doped" in order to search for *d*-wave superfluidity. As we approach that benchmark, new techniques will be required with a number of solutions already proposed [200-202]. An alternative, which is less demanding on resolution but works only in certain limits, is to look at the width of the in-situ distribution [58,75,107].

Another method related to in-situ imaging is to measure the distribution of in-situ site occupancies. Determining temperature requires measuring the ratio of the atoms that reside on sites occupied by *n* atoms. The most straightforward approach is to use site-resolved imaging, discussed above, to measure the local site occupancies. This allows the distributions to be measured in a single phase of the system, without needing to average over the trap. In Ref. [36] temperature was determined by measuring the distribution in both the $n=1$ and $n=2$ Mott-insulator regions of a 2D boson lattice system (Figure 15). The variance around 1 and 2 atoms per site was fit to a QMC curve to extract temperature (with about 20% error), although analytic forms are available in the zero-tunneling limit.

High-resolution imaging is not a requirement, however, for measuring in-situ number distributions. There are two primary methods for measuring the number of sites with *n* atoms that do not rely on high resolution imaging, both of which measure the number of atoms transferred into an initially unpopulated state. In the first technique, spin changing collisions between pairs of atoms on a site cause population to shift into other spin states [203]. The other process uses an external field to transfer atoms into another state; interaction effects, which depend on the number of atoms per site, affect the transfer. The occupation statistics can be measured by investigating resonances that occur at different frequencies of the external field[64,116,204].

While number distributions have been analyzed for bosonic [65,203,205-207] and fermionic systems[116,204], this method has been extensively evaluated, specifically for thermometry, in fermionic systems where the number of atoms per site is limited to zero, one, and two [116,204]. The number of doubly occupied sites as a function of temperature in the Fermi-Hubbard model can be calculated from theory with relative accuracy in certain limits [116,184,204,208]. A comparison of QMC, DMFT, and series approximations demonstrated that simple series approximations give almost identical results to the more complex computational calculations in the parameter range of current experiments [116]. The most important issue with this thermometry method is that there are certain regimes in which double occupation is not a sensitive function of temperature. A detailed study of using double occupation to measure the in-lattice entropy achieved results with less than 25% uncertainty [116]. Also, this method is only useful above $S/N \approx k_B$, because number fluctuations are absent at lower entropy.

## 4.4. Probing using light

Another approach to thermometry in lattices is to probe the system using light. If the light is not resonant with an electronic transition, it will scatter from the atomic gas, which acts as a spatially dependent refractive medium. Because the scattering is coherent, there will be interference between the light scattered from different lattice sites, therefore making it a probe of the density-density correlations in the lattice [209-216]. To sensitively probe thermal effects, one possibility is to block the elastically scattered portion of the light and measure the total amount of inelastically scattered power [216]. The total amount of light detected in this scheme depends on temperature; technical noise is a concern because the amount of light scattered is small. For example, at $T/T_F = 0.1$, one hundred experimental runs are required for the uncertainty in temperature arising from photodetector shot noise to be driven below 20%. A variation on this method is to place the detector at an angle that does not satisfy the Bragg condition for elastic scattering (if there were identical numbers of particles in each site, e.g., such as in the MI phase). The number of photons scattered into this angle depends on the temperature and the phase of the system [215]. A number of proposals have also considered using a cavity to detect the scattered light [212-214].

By choosing suitable parameters of the light (such as polarization and frequency), the atomic refractive index can be spin dependent. Therefore, light scattering is one method for detecting spin correlations (such as the AFM state). One proposal [217] is to direct the beam at a specific angle so that elastic scattering only occurs when the system is spin ordered (Figure 19). A variation of this is to filter based on polarization [210], so that only light which scatters into an orthogonal polarization is detected, thereby allowing detection of the light that is coupled to atomic spin operators. Spin correlations also appear as a polarization rotation (through the Faraday effect) in the beam that is not scattered [209]. High resolution can be obtained if the probe light is also retro-reflected and forms a lattice. In this scheme, different spin phases can be identified by measuring the rotation of the probe polarization as the probe lattice position relative to the atomic lattice is varied. Spin thermometry is possible by measuring the degree of spin correlations, by, for example, determining the fraction of the diffracted power. Above the Neel temperature, however, these approaches provide no signal.

An alternative approach is to measure the effect of the light on the atoms, either through Bragg or Raman scattering. In this scheme, the system is excited using two laser beams with a photon energy and momentum difference determined by the frequency difference and angle between the beams, respectively. Several different definitions for light scattering processes have been used in the literature; here, we will call this process "Bragg scattering" if the atoms change only their motion, and "Raman scattering" if the internal state changes. For either, excitations in the gas are created if the energy and momentum difference of the beams corresponds to a transition to an excited state. Internal state dynamics, such as changes in the interaction strength, may play a role in Raman scattering. Excitations created by light scattering can be measured as an increase in the size of the gas after TOF or, for Raman scattering, by counting the atoms transferred to a different hyperfine state.

Bragg scattering is not a sensitive measure of temperature in the superfluid regime, but in the MI regime could be used to measure temperatures on the order of $U/5$ [218]. In the MI regime finite temperature causes peaks to appear in the spectrum which do not occur at zero temperature. The appearance of these peaks has been observed in 1D [98], but more theory is required to relate the observed signal to a temperature. The temperature dependence of the Raman signal has been calculated theoretically, but is it strongly model dependent [219].

### 4.5. Extrinsic thermometer

The final strategy for thermometry is to thermally connect the lattice gas to an auxiliary system or degree of freedom that can be used as a primary thermometer. The advantage of this method is that measurable properties of the thermometer system can have a simple analytic dependence on temperature. One implementation of this method is to use the spin degree of freedom as the thermometer for the motion. Spin ordering is expected at very low temperatures in the Hubbard models; above these temperatures, however, the spins on different lattice sites are essentially decoupled from one another. Therefore, the atomic distribution, which is the degree of freedom we are interested in, is decoupled from the spin degree of freedom (as long as the strength of the interactions are not strongly state dependent). By adding a spatially-dependent energy scale for the spins, in the form of a magnetic field gradient, one can extract the spin temperature based on the width of the domain wall. This technique was demonstrated experimentally in Ref. [107] where temperatures as low as 1 nK were measured, with the lower bound set by imaging resolution and the superexchange temperature. One concern is that poor coupling between spin and particle degrees of freedom may mean that the spin temperature from the domain wall does not reflect the temperature at the edge of the gas [165]. While the spin temperature was compared to the particle temperature in the high temperature limit, there has been no independent confirmation at low temperatures.

Another proposal is to use a different atom that does not experience the lattice potential as a thermometer. This can be accomplished by polarizing the lattice beams in a specific fashion [47] and using a different spin state, or by using a different species altogether and a "magic" wavelength for the lattice light [157,159]. In this way, thermometry can be performed on the atoms that are only harmonically trapped using the traditional methods. The main limitation to these methods is whether or not the thermometer is truly in thermal equilibrium with the system under study. Interactions will maintain thermal contact, but the timescale for thermalization may be too long compared to loss processes and heating in the system. Another limitation is that the presence of the thermometer may perturb the system and visa-versa. For example, if the presence of the lattice atoms significantly alters the potential felt by the harmonically trapped atoms, extracting temperature using the traditional methods will no longer be possible and the thermometer will not be effective. Or, along the lines of self-heating in experiments on solids, adding the thermometer atoms may introduce new technical heating sources, such as inelastic collisions. Also, the heat capacity of the harmonically trapped species will introduce limitations [47].

### 4.6. Conclusion

In this section we have reviewed the five main strategies for thermometry in optical lattices: entropy matching, analyzing the momentum distribution (TOF imaging), in-situ measurement, scattering with light, and using an extrinsic thermometer. While all have their benefits and disadvantages, we would like to address the question of the most viable strategy for developing a technique that can reach into the lowest entropy regimes. While the most mature techniques are entropy matching and TOF imaging, these cannot be the primary methods in the future. Both require extensive theoretical input, and, in the case of entropy matching, rely on the assumption of adiabaticity, which will not apply as we pursue lower entropy phases. Light scattering is a promising approach, yet it has not been experimentally demonstrated and requires specific theoretical models. The opposite is true for using a thermometer system, which is model independent. Yet, such a thermometer has not been satisfactorily demonstrated and important questions have been raised about thermal contact between lattice gases and external degrees of freedom. While these questions should be pursued, they may limit the viability of this approach.

This leaves in-situ measurement, which we feel may be the principal technique for measuring temperature in future lattice experiments. Recently, several groups have demonstrated high-resolution imaging in a lattice [36,66,67,139,189], thereby surpassing the main hurdle of in-situ imaging. Since the technical challenges have been overcome, there are clear advantages of in-situ imaging; we will highlight two. First, local probes do not have to average over several in-trap phases, which vastly clarifies interpretation of data. Second, in deep lattices, the Hamiltonian is nearly diagonal in the number basis. With little theoretical input, temperature can be measured from distributions [67,139] and local number fluctuations [36]. These methods are viable over a large temperature range, until density fluctuations are frozen out and magnetic ordering sets in. Although a main challenge is *spin dependent* in-situ imaging, the resolution has been achieved [36,67] to in principle detect local magnetic ordering, which will naturally extend temperature measurements into interesting magnetic phases. Although a few challenges remain—spin-resolved imaging, 3D systems, for example—in-situ imaging is the most promising technique for thermometry in a lattice for low entropy phases.

## 5. Conclusion

In summary, optical lattices are well suited to contribute to our understanding of a wide variety of models of strongly correlated materials. Unfortunately, energy and temperature scales for Hubbard models as realized by optical lattices are extremely small. Quantum phases with energy scales such as $t$ and $U$ are within the range of current cooling techniques, and accordingly have been realized in numerous experiments. However, the next lowest possible energy scale $J$ (super-exchange) has not been reached. We must cross this barrier to explore magnetic phases and ultimately search for $d$-wave superfluidity. In this review, we addressed the issues involved with cooling to and below the temperature scale associated with magnetic ordering. Our main conclusions are:

(1) The standard technique for cooling in optical lattice experiments—evaporative cooling in a harmonic trap followed by slowly turning on the lattice potential—will likely not reach sufficiently low entropy per particle. New lattice cooling techniques must be developed and investigated experimentally.

(2) Developing new cooling techniques will require a detailed understanding of heating processes in the lattice, since the cooling power must exceed the heating rate in the regime of interest. From a single particle analysis of lattice-light-induced heating, we conclude that it is *in principle* possible to achieve magnetic ordering for large lattice laser wavelengths. However, a complete study will require a many-body quantum mechanical description of light scattering in a densely populated optical lattice.

(3) We lack a complete theory of dynamics and thermalization in strongly correlated lattices. In particular, we do not understand the exchange of entropy between the spin and motional degrees of freedom. Evaluating cooling proposals without such a theory is impossible, since the feasibility of any method strongly depends on competing timescales.

(4) New thermometry methods are required to develop and evaluate cooling techniques. We feel that the principle approach for future experiments will be thermometry based on site-resolved imaging.

Reaching the super-exchange temperature will require a two-pronged approach. On the one hand, we need to continue to develop and investigate new cooling and thermometry techniques. A number have already been proposed and tested, which are summarized in Figure 21. Simultaneously, the physics of heating and thermalization must be explored. Vigorous activity on both fronts is underway, making a new regime of lattice physics a tantalizing prospect.

## 6. Heating and Dipole Potential Appendix

In this appendix, we discuss in further detail the light-induced energy dissipation rates and dipole potential for the three common lattice configurations shown in Figure 20. Following Ref. [125], the energy dissipation rate for a two-level atom in a light field is, in general,

$$\dot{E} = E_R \frac{1}{8}\frac{\Gamma^3}{\Delta^2}\frac{I(\vec{x})}{I_{sat}}\left[1+\frac{|\vec{\nabla}E(\vec{x})|^2}{k^2|\vec{E}(\vec{x})|^2}\right].$$

Here $\vec{E}(\vec{x})$ is the electric field, $I(\vec{x}) = c\varepsilon_0|\vec{E}(x)|^2/2$ is the optical intensity, $I_{sat}$ is the saturation intensity, we work in the rotating frame and ignore counter-rotating terms, and we assume that the light is far from resonance. In the last section of this appendix, we introduce the contribution from the counter-rotating term, which is important at large detunings (comparable to the transition frequency).

## 6.1. Counter-Propagating Lattice

For a lattice formed from counter-propagating laser beams (Figure 20a), $E(\vec{x}) \propto \cos(kx)$, and $\dot{E} = E_R\frac{1}{8}\frac{\Gamma^3}{\Delta^2}\frac{I_0}{I_{sat}}$, where $I_0$ is the intensity at a standing-wave anti-node. Using $V_{lat} = \frac{\hbar}{8}\frac{\Gamma^2}{\Delta}\frac{I_0}{I_{sat}}$ for the lattice potential depth, the energy dissipation rate can be conveniently expressed as $\dot{E} = E_R\frac{V_{lat}}{\hbar}\frac{\Gamma}{|\Delta|}$, which is independent of position in the standing wave. The heating rate is identical for spin-dependent "lin-$\theta$-lin" lattices [47], where $V_{lat}$ is the potential depth of the lattice when $\theta = 0$. When $\theta = \pi/2$ and the lattice is completely spin-dependent, the heating rate is independent of state and field direction.

## 6.2. Lattice beams intersecting at an angle

Optical lattices are often formed using laser beams that intersect at an angle, as in Figure 20b. Referring to that geometry shown in the figure, $\vec{E}(\vec{x}) \propto e^{-iky\sin(\theta)}\cos(kx\cos\theta)$, leading to a standing wave $|\vec{E}(\vec{x})|^2 \propto \cos^2(kx\cos\theta)$ with spacing $d = \lambda/2\cos(\theta)$ between wells. The energy dissipation rate is $\dot{E} = E_R\frac{V_{lat}}{\hbar}\frac{\Gamma}{|\Delta|}[(1+\sin^2\theta)\cos^2(kx\cos\theta) + \cos^2\theta\sin^2(kx\cos\theta)]$, which is spatially dependent. To compare heating rates between lattices made using different geometries, it is useful to rewrite this as:

$$\dot{E} = \frac{\hbar^2}{m}\frac{V_{lat}}{\hbar}\frac{\Gamma}{|\Delta|}\frac{\pi^2}{2d^2}[(1+2\tan^2\theta)\cos^2(kx\cos\theta) + \sin^2(kx\cos\theta)].$$

Although the heating is spatially dependent, the atoms only reside in the nodes (anti-nodes) for a red (blue)-detuned lattice, and so we consider only these two cases. For red-detuned lattices, the atoms are localized at $x \approx 0$ and so,

$$\dot{E} = \frac{\hbar^2}{m}\frac{V_{lat}}{\hbar}\frac{\Gamma}{|\Delta|}\frac{\pi^2}{2d^2}\left(\frac{8d^2}{\lambda^2} - 1\right),$$

where $\lambda < 2d$. Therefore, for lattices with equal well spacing and potential depth, the heating rate is lowest using the longest wavelength possible, which corresponds to the retro-reflected configuration, i.e., the results of Section 2.3 are not improved by using an angled lattice.

For blue-detuned lattices, the atoms are located at $x \approx \frac{\pi}{2k\cos\theta}$ and so,

$$\dot{E} = \frac{\hbar^2}{m} \frac{V_{lat}}{\hbar} \frac{\Gamma}{|\Delta|} \frac{\pi^2}{2d^2}.$$

In this case, for lattices with equal well spacing and potential depth, the heating rate is lower for shorter wavelengths (i.e., larger detunings). However, this argument does not apply to arbitrarily short wavelengths, since for large detunings this formula must be corrected by a pre-factor $(\lambda_0/\lambda)^3$ (see Section 6.5). Once the lattice is detuned by hundreds of nanometers (the regime in which blue-detuned lattices achieve the best figures-of-merit, see Figure 12), the pre-factor outweighs any benefit of using a shorter wavelength, angled lattice.

### 6.3. Lattice Beams with Different Intensities

An important issue is the effect of imperfect lattice contrast on the heating rate. It is common in lattice experiments employing a counter-propagating geometry to have, by necessity, unequal intensity in the two laser fields (Figure 20c). In this case, $\vec{E}(\vec{x}) \propto \left[\cos(kx) - \alpha e^{-ikx}\right]/\sqrt{1-2\alpha}$, with $(1-2\alpha)^2 = I_2/I_1$ as the ratio of intensities. A standing wave $|\vec{E}(\vec{x})|^2 \propto \left[\cos^2(kx) + \alpha^2/(1-2\alpha)\right]$ is generated with minima that are not completely dark. Once again, the dissipation rate $\dot{E} = E_R \frac{V_{lat}}{\hbar} \frac{\Gamma}{|\Delta|}\left[1 + \frac{2\alpha^2}{1-2\alpha}\right]$ is independent of position, but higher than a "perfect" lattice with the same lattice potential depth. Fortunately, $\dot{E}$ increases very slowly as the ratio of laser intensities deviates from unity, reaching only 1.06 times the "perfect" lattice value for $I_2/I_1 = 0.5$, for example.

### 6.4. Beyond Two Levels

When the detuning is comparable to the fine structure splitting between the $P_{3/2}$ and $P_{1/2}$ states we must consider these two levels independently for heating (the effect on the potential is already listed in the main text). Then the heating rate is,

$$\frac{\dot{E}}{E_R} = \frac{I_0}{24}\left(\frac{\Gamma_{3/2}^3}{\Delta^2_{3/2} I_{sat\,3/2}} + \frac{\Gamma_{1/2}^3}{\Delta^2_{1/2} I_{sat\,1/2}}\right)$$

The total heating rate due to all transitions is the sum of the heating rates from the individual transitions, provided that $I/I_{sat} \ll 1$.

### 6.5. Large Detuning Lattices

In all the previous expressions in this appendix we have assumed that, for simplicity, the detuning is small compared to the transition frequency. However, since the discussions in the text have motivated the need for large detunings, we must consider going beyond this limit.

New terms ("counter-rotating") terms) become important and must be included in the equations for the dipole force and heating rate.

The dipole potential for a two-level atom is:

$$U_{dip} = -\frac{3\pi c^2}{2\omega_0^3}\Gamma\left(\frac{1}{\omega_0-\omega_L}+\frac{1}{\omega_0+\omega_L}\right)I(\vec{x}),$$

which is correct for all detunings much greater than the natural linewidth [46].

Also, at large detunings the *decay rate* must be corrected by a pre-factor of $(\omega_L/\omega_0)^3$ to account for the changing blackbody density of states. Therefore the large detuning heating rate (for a two-level atom) is

$$\dot{E} = E_R\frac{3\pi c^2}{2\hbar\omega_0^3}\Gamma^2\left(\frac{\omega_L}{\omega_0}\right)^3\left(\frac{1}{\omega_0-\omega_L}+\frac{1}{\omega_0+\omega_L}\right)^2 I(\vec{x})$$

where $I(x)$ is the maximum intensity in the lattice.

## 7. Acknowledgements

The authors acknowledge support from the DARPA OLE program, the Office of Naval Research, and the National Science Foundation. The authors also wish to thank Andrew Daley and Joseph Thywissen for helpful conversations.

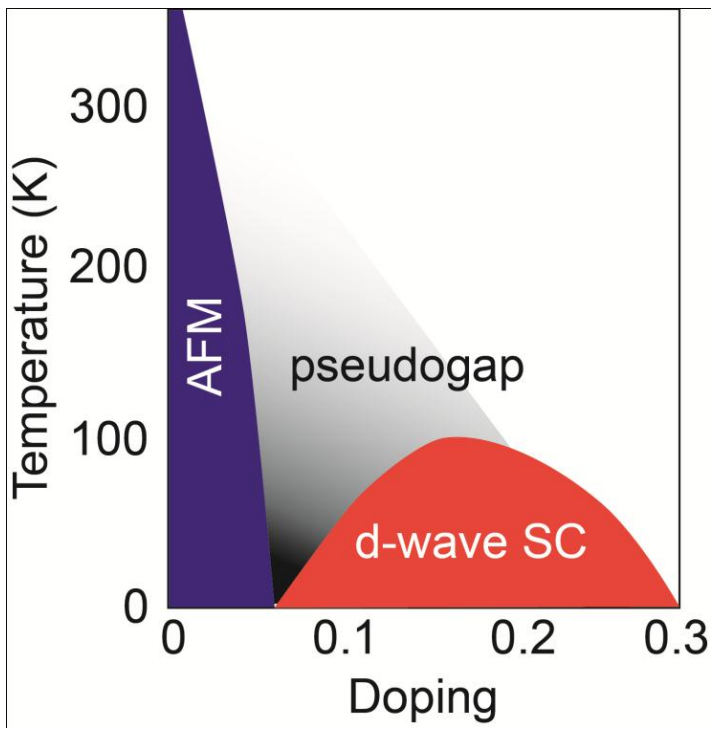

**Figure 1. Simplified cuprate phase diagram.** The phases present as temperature and charge doping (per copper atom away from half-filling) are varied are shown.

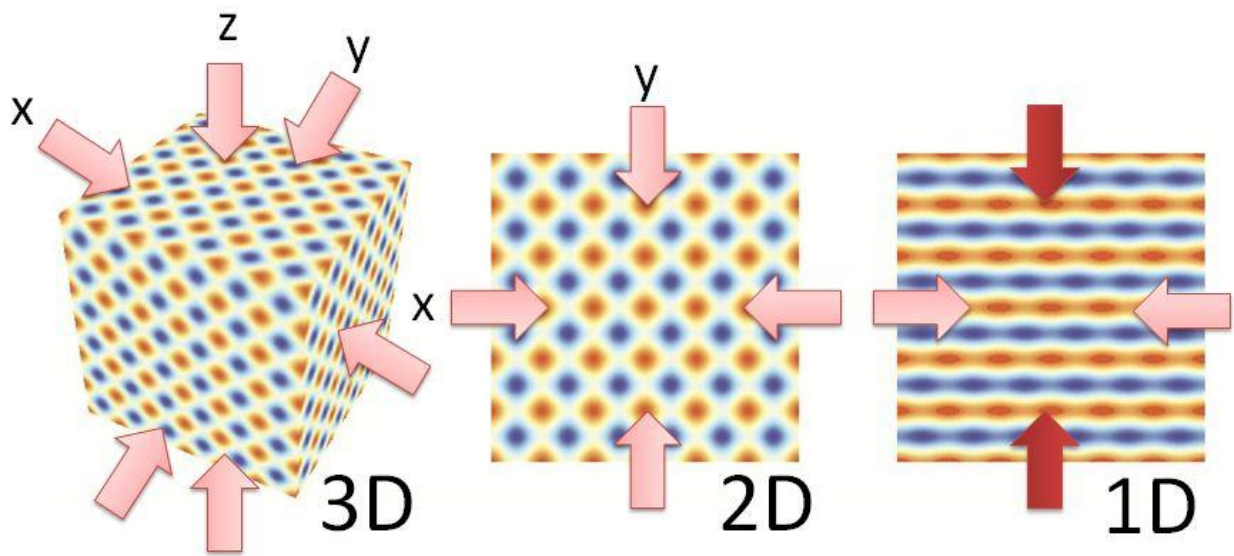

**Figure 2. Optical lattices based on a cubic geometry.** The lattice potential is shown in false colour. To create a 2D lattice, the atoms are confined into "pancakes" along the $z$ direction by increasing the power in one pair of lattice beams. The power is increased in the $z$ and $y$ beams to confine the atoms to a series of tubes and create a 1D lattice.

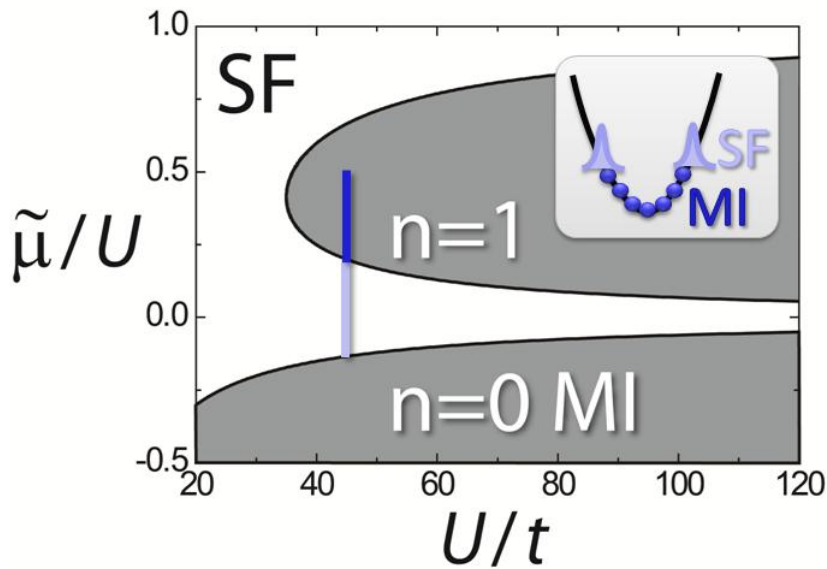

**Figure 3.** Zero-temperature Bose-Hubbard phase diagram for a homogenous system. The trapped system can be treated using the local-density approximation and introducing an effective chemical potential $\tilde{\mu}$. The atoms are then understood as sampling a vertical line on this phase diagram, with the edge of the gas located at zero filling and the center terminating at a maximum $\tilde{\mu}$. The inset shows the inhomogeneous configuration of phases present for the line displayed in the phase diagram.

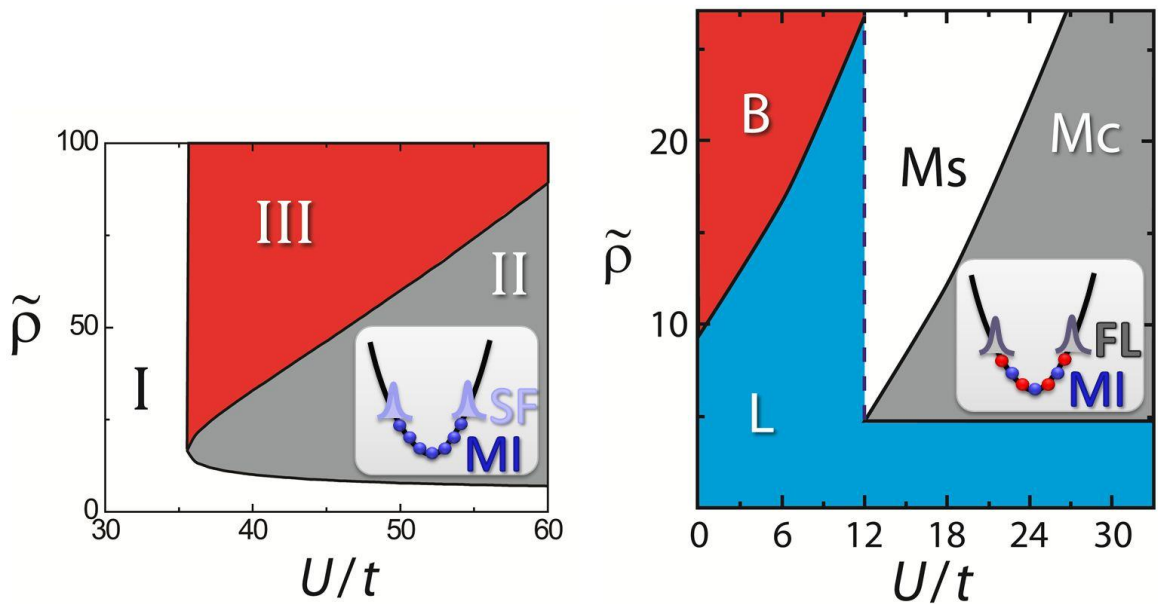

**Figure 4.** Zero-temperature phase diagram for trapped Bose-Hubbard (left) and high-temperature phase diagram for trapped Fermi-Hubbard (right) gases. Three phases are shown for bosons: the entire gas is a superfluid (I), a Mott insulator core surrounded by a superfluid shell (II), and a nested superfluid–Mott insulator–superfluid phase (III). For fermions, Fermi-liquid (L), band-insulator (B), Mott insulator core surrounded by Fermi liquid shell (Mc), and metallic core surrounded by Mott insulator shell (Ms) phases are shown. The insets show the phases (II and Mc) we consider for addressing the impact of light-induced heating in Section 2.3.

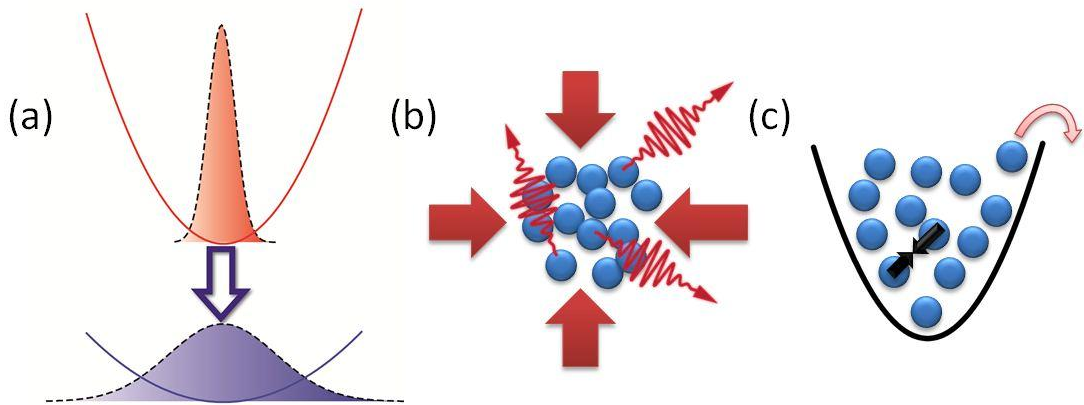

Figure 5. Cooling methods. (a) Cooling via adiabatic expansion: the gas cools as the confinement is adiabatically relaxed. No entropy is removed, even though the temperature can be drastically reduced. Therefore, low entropy states such as an anti-ferromagnetic ordering or *d*-wave superfluidity cannot be accessed using this method. The predominant cooling methods used in ultra-cold atom gas experiments are shown in (b) laser cooling and (c) evaporative cooling. Entropy is removed by each process, in contrast to adiabatic expansion. Entropy is carried away by scattered light in laser cooling and by high energy atoms ejected from the trap in evaporative cooling.

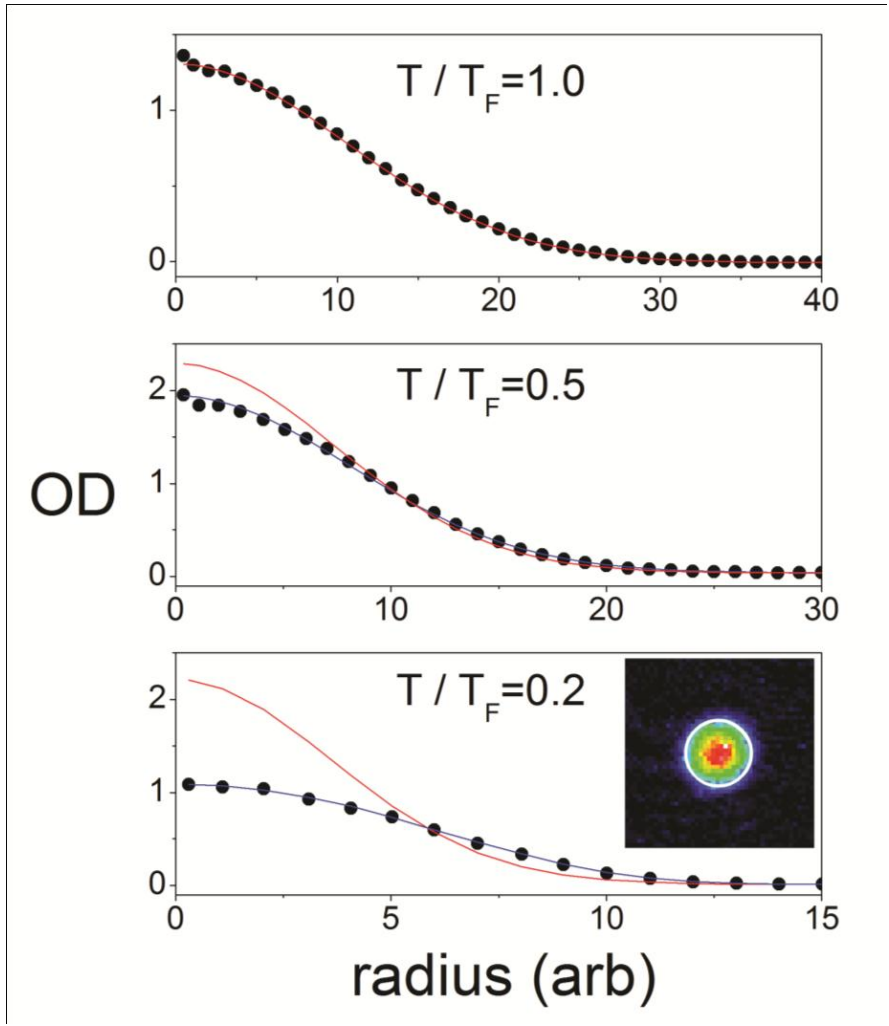

Figure 6. Data showing TOF thermometry for an ideal Fermi gas. The inset shows a TOF image taken of a gas of $^{40}$K atoms that have been cooled to approximately a fifth of the Fermi temperature, and therefore most of the atoms have a momentum below the Fermi momentum (indicated by the white line). The optical depth (OD)—proportional to the density—is plotted for such TOF images that have been angularly averaged. The profile is fit to a hypergeometric function (blue) to determine temperature. Temperature can also be extracted from a Gaussian fit (red) to the low signal-to-noise ratio tail of the distribution.

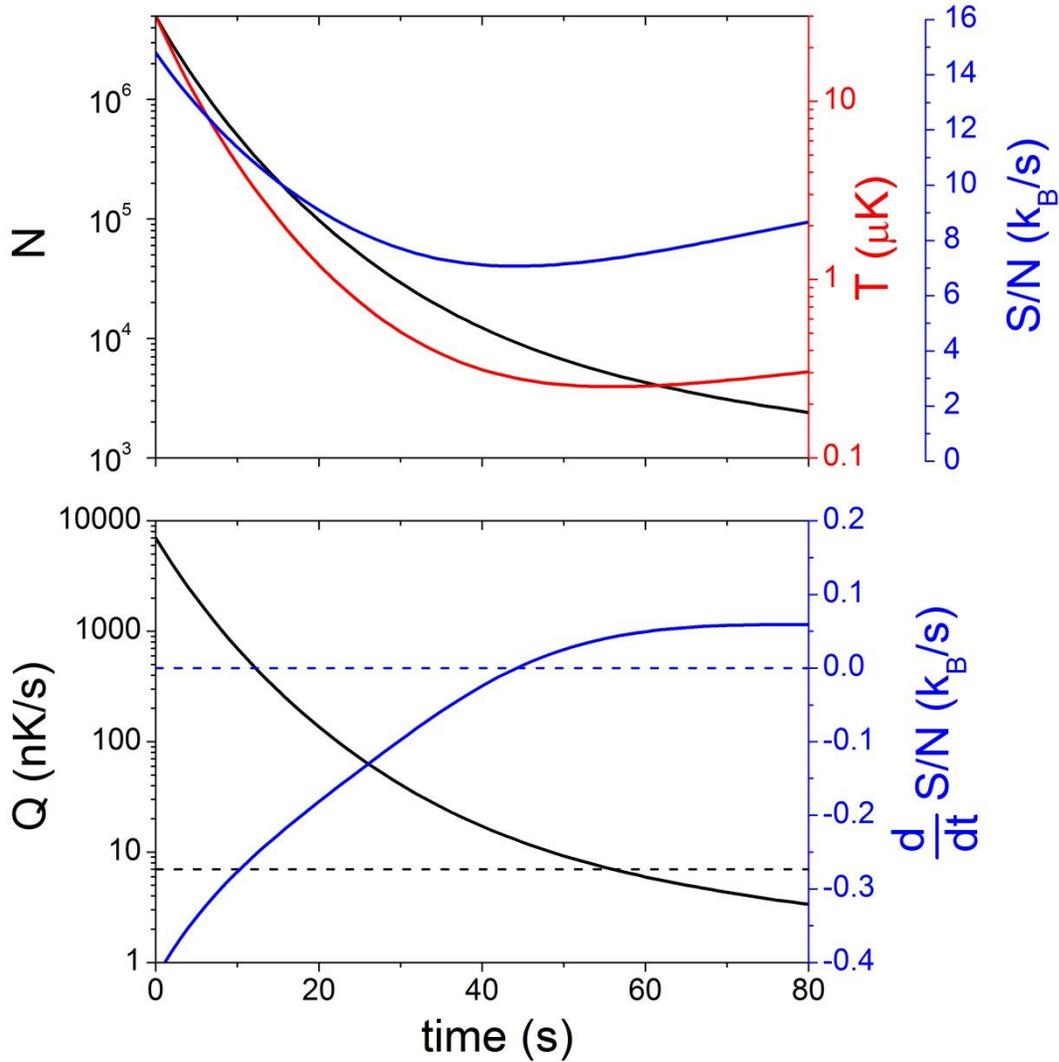

Figure 7. Evaporation trajectory (top) and cooling power (bottom). In the top graph, the temperature (red), number of atoms (black), and entropy per particle (blue) are shown as the evaporation proceeds in time. For this simulation we start with 5 million atoms at 30 μK in a trap with $\omega = 2\pi 100$ Hz, and we have assumed a constant heating rate $\dot{T}_{heat} = 7\ nK/$s nK/s, $\eta = 5$, and loss time constant $\tau = 100\ s$ (where $\dot{N}/N = -1/\tau$). In an experiment, the best choice for $\eta$ results in the smallest change in the number of atoms $N$ for the largest change in temperature $T$. Typically, the optimal $\eta$ is between 3 and 5 and is determined by the ratio of elastic rethermalizing collisions to the loss rate induced by inelastic collisions and collisions with residual gas atoms. In the bottom graph, the cooling power (black) and rate of change of entropy (blue) are shown. The dashed lines represent the heating rate (black) and the boundary between heating and cooling (blue). Absolute temperature starts to increase when the cooling power drops below the heating rate, at about 55 s. Cooling (in the sense of reducing entropy per particle) fails when the rate of change of entropy becomes positive slightly earlier at 45 s.

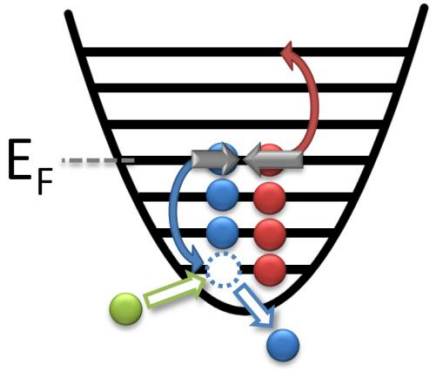

Figure 8. Hole heating in a Fermi gas. Residual gas atoms (green) remove atoms from low energy states in the trap through collisions, leading to an increase in entropy per particle. Elastic collisions between two spin species (red and blue) then re-populate the empty state.

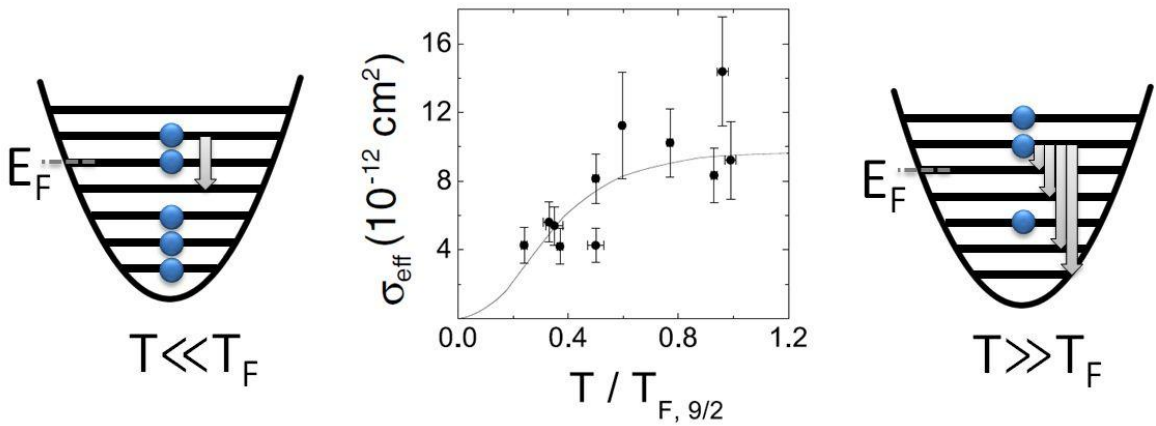

Figure 9. Pauli blocking of collisions. Elastic collisions necessary for rethermalization are suppressed by the Pauli exclusion principle in the quantum degenerate regime, when most states below the Fermi energy are occupied. The arrows indicate transitions to possible final states following a collision for an atom with energy slightly above the Fermi energy. The effective cross-section for rethermalizing collisions is reduced in the quantum degenerate regime, as shown in the data (reproduced from Ref. [123], copyright 2001 by the American Physical Society) from measurements using a gas of harmonically confined $^{40}$K atoms.

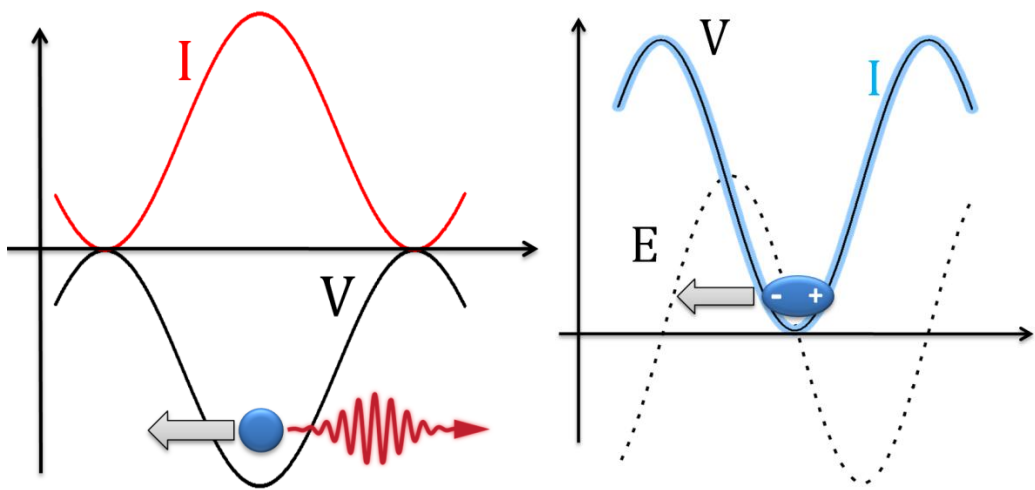

Figure 10. Heating arising in an optical lattice from photon recoil (left) and fluctuations in the atomic dipole (right). The minimum in the potential $V$ (black) occurs at a maximum in the laser intensity $I$ (red) for a red-detuned lattice (left). Heating arises from the random recoil (gray arrow) following photon scattering events. In a blue-detuned lattice, the potential $V$ follows the optical intensity $I$ (right). The gradient of the magnitude of the electric field $E$ has a maximum at the minimum in the potential. The electromagnetic vacuum induces a fluctuating atomic dipole, which then experiences a force (grey arrow), leading to heating.

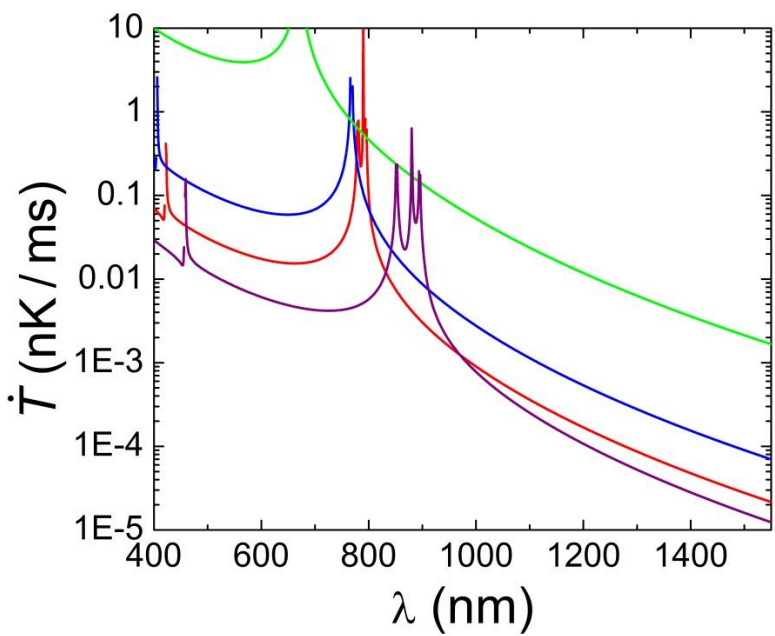

Figure 11. Absolute heating rate for the lattice conditions considered in this section. The colour scheme, used for all figures in this section, is: $^{133}$Cs (purple), $^{87}$Rb (red), $^{40}$K (blue) and $^{6}$Li (green). For this and the remaining figures in this section, we include contributions from the D1 and D2 transitions, but we make several approximations that are only correct for detunings that are large compared with the natural linewidth (roughly 6 MHz for the D1 and D2 transitions) and ground-state hyperfine splitting (e.g., approximately 10 GHz for $^{133}$Cs). We also include the $4s \rightarrow 5p$, $5s \rightarrow 6p$, $6s \rightarrow 7p$ and transitions for $^{40}$K, $^{87}$Rb, and $^{133}$Cs, respectively. For the discussion in the text, however, we use a simple two-level model for the atom for clarity. Under realistic experimental conditions, sufficient optical power is available to cover the wavelength range under consideration: appropriate high power (>10 W) solid state and fiber lasers lasers are available at 532, 1064, and 1550 nm, and over 1 W can be produced across the 600–900 nm range using dye and Ti:sapphire lasers. Achieving the power necessary to create a lattice at 400 nm with the desired parameters is feasible for all but Li using frequency doubling [98]. Although laser intensities can be high in a lattice, it is unlikely that photoionization will have a detrimental effect for short wavelength lattices, since the laser wavelength can be tuned away from exciting intermediate electronic states [99,100].

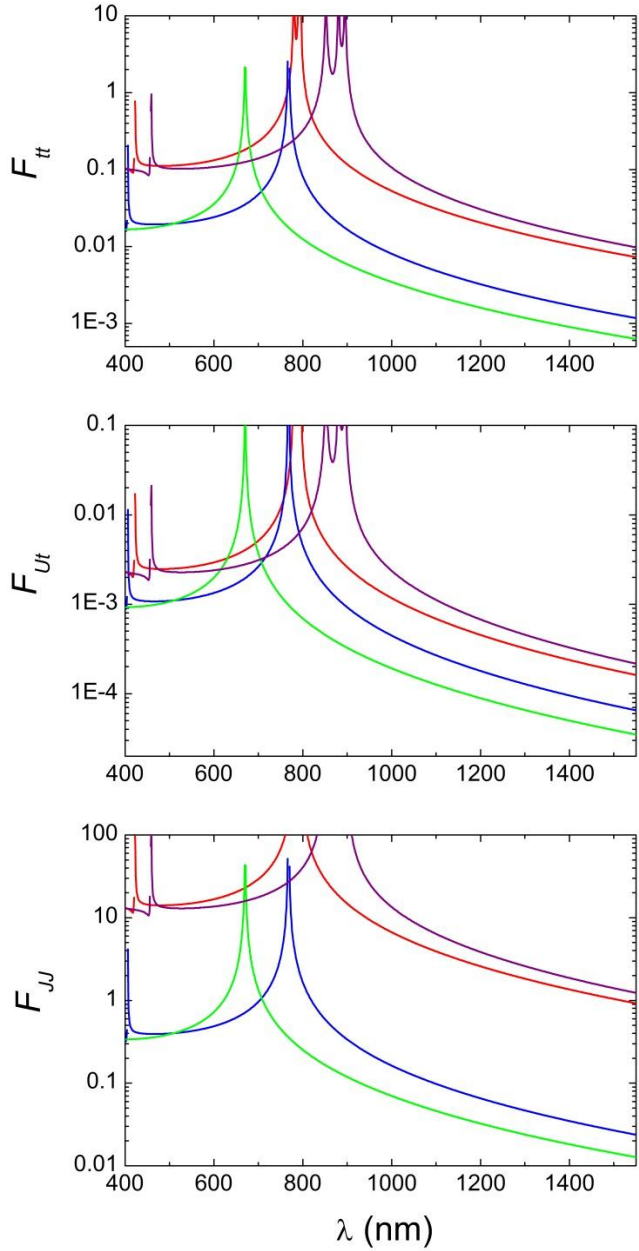

Figure 12. Figures of merit for cooling to low entropy phases. In this section, we fix the characteristic density by adjusting $N$ and $\omega$ within reasonable experimental bounds. Keeping $U/t$ constant is accomplished by controlling the lattice potential depth $s$ according to $s = \left[ ln^2 \left( U\lambda / \pi\sqrt{2} a_s t \right) \right]/2$, which gives $s \approx 11$ and $s \approx 14$ for fermions and bosons, respectively.

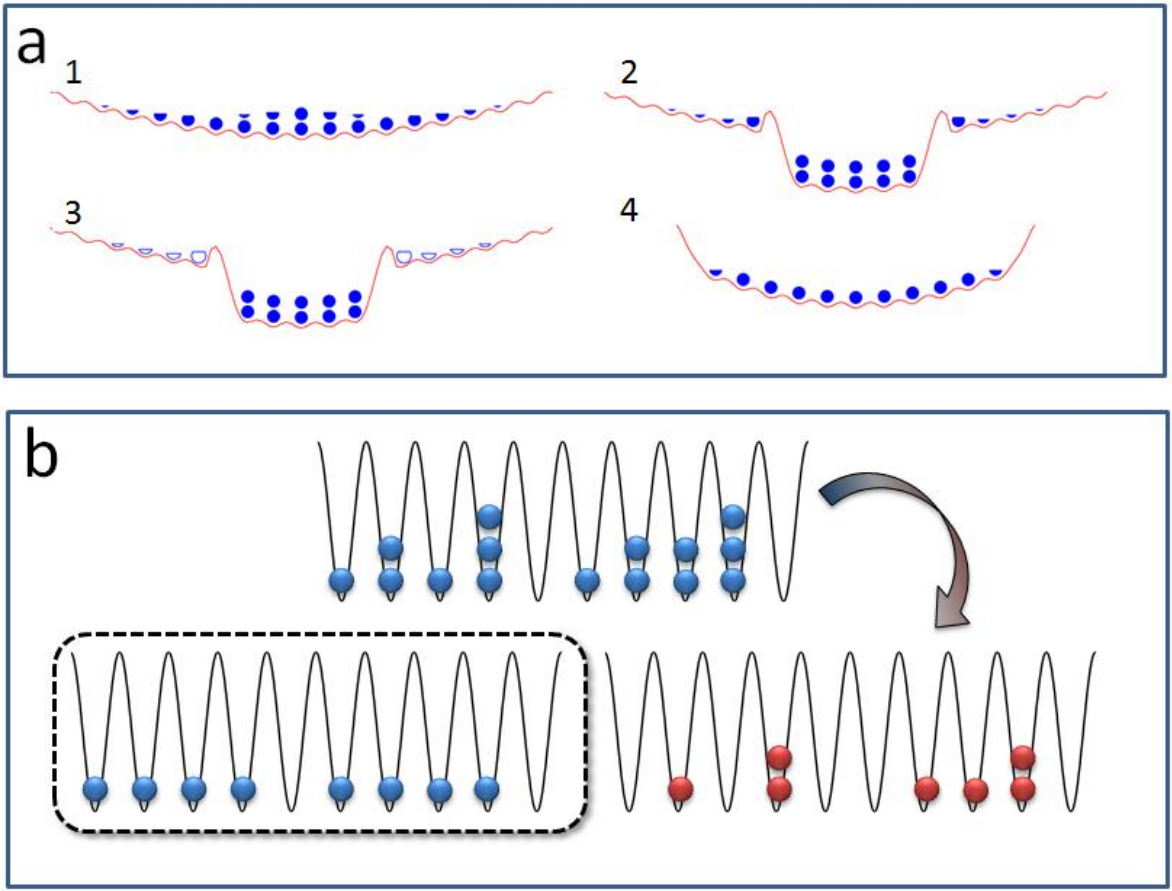

Figure 13. Filter cooling. In one method shown in (a), entropy is removed by shaping the confinement (adapted with permission from Ref. [143], copyright 2009 by the American Physical Society). Number filtering schemes (b) transfer entropy into an auxiliary state (red atoms); these atoms are then ejected from the lattice. A limitation of this scheme is that initially empty sites cannot be removed from the final state.

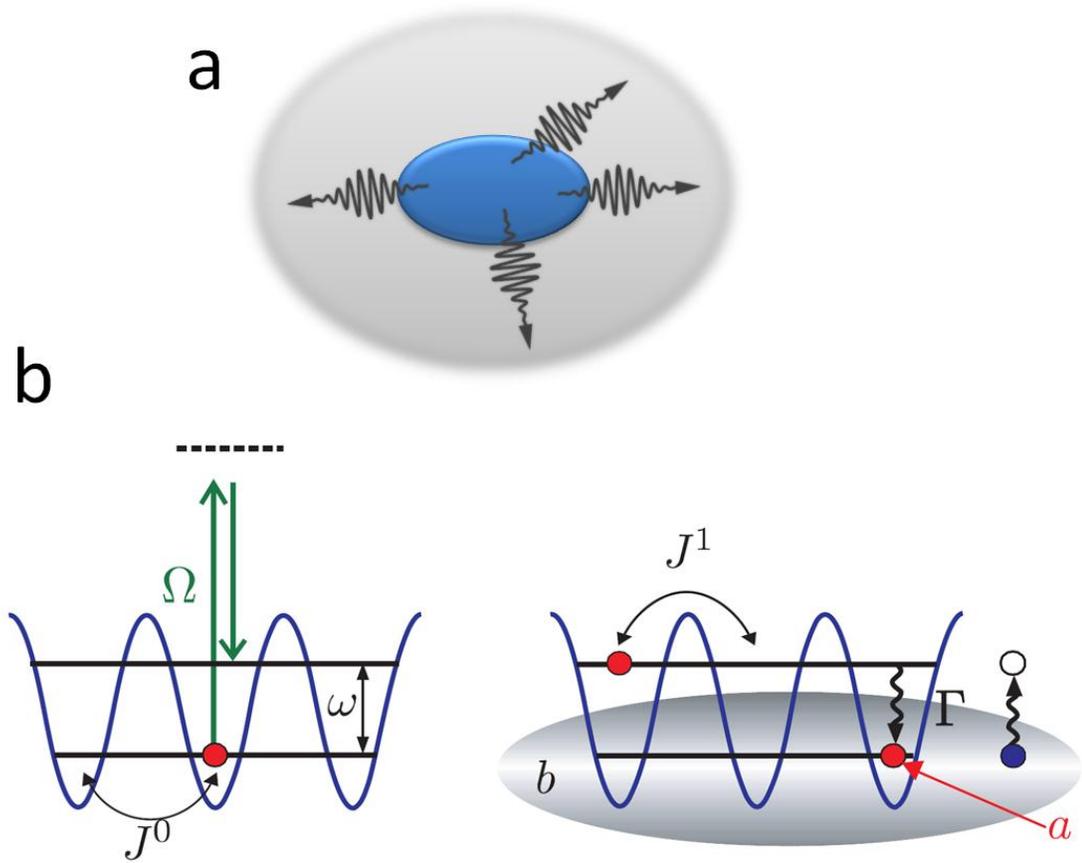

Figure 14. Immersion cooling. Schematically (a), these cooling methods work by driving entropy from the system (blue) into an entropy reservoir (gray). One immersion cooling scheme shown schematically in (b) drives atoms into excited bands (left). Decay into the ground band proceeds by release of energy and entropy into a co-trapped BEC (right). Cooling is achieved by selectively driving the high energy atoms into excited bands and eventual accumulation after decay into a "dark state" near zero quasimomentum in the ground band (reproduced from Ref. [168]).

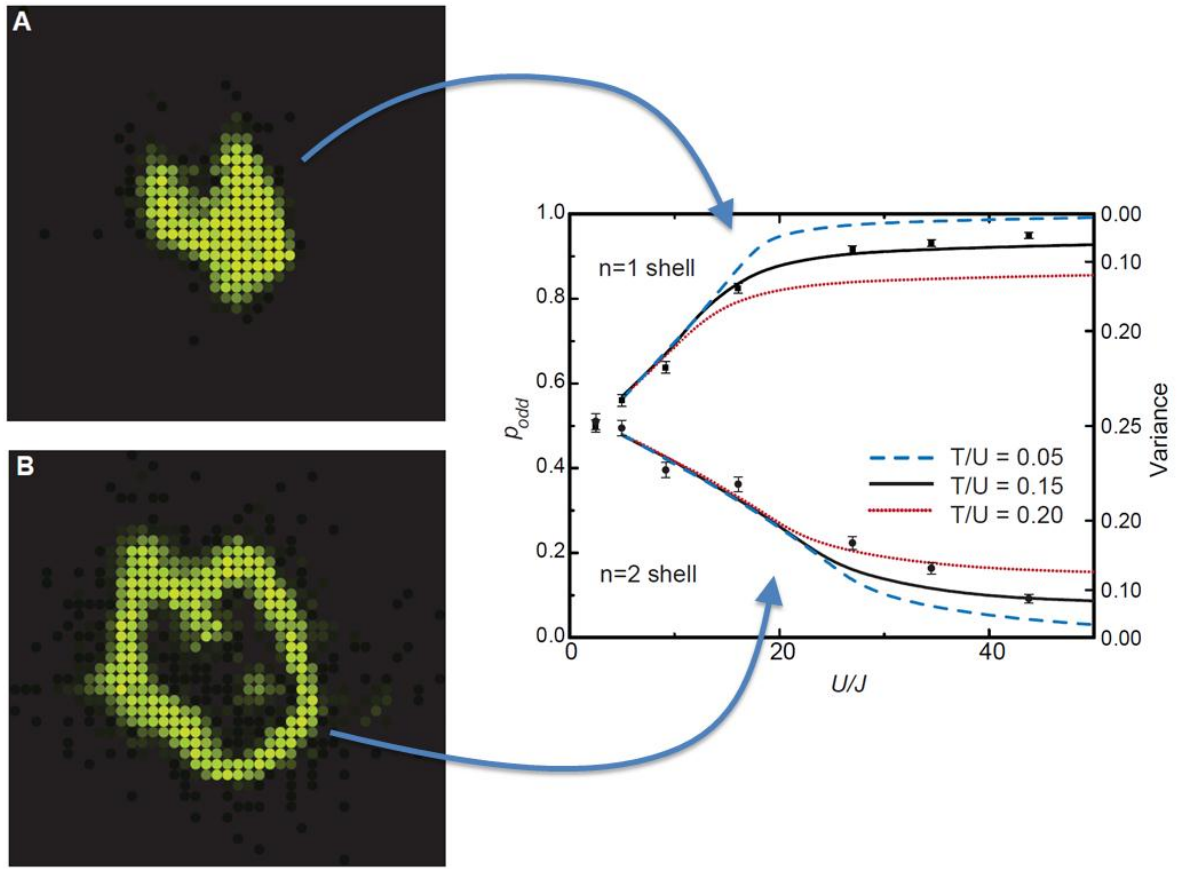

Figure 15. In-situ thermometry via fluctuations. In-situ images (left) of the MI state averaged over 20 experimental runs. The imaging process returns the number of atoms in each site modulo 2 due to rapid light-assisted collisional loss. The top image shows a unit filling MI phase, and the bottom image an (invisible) MI phase with two atoms per site (surrounded by a visible SF shell). The number statistics in the MI phase (right) is fit to QMC calculations to determine temperature. All images were reproduced from Ref. [36]. Reprinted with permission from AAAS.

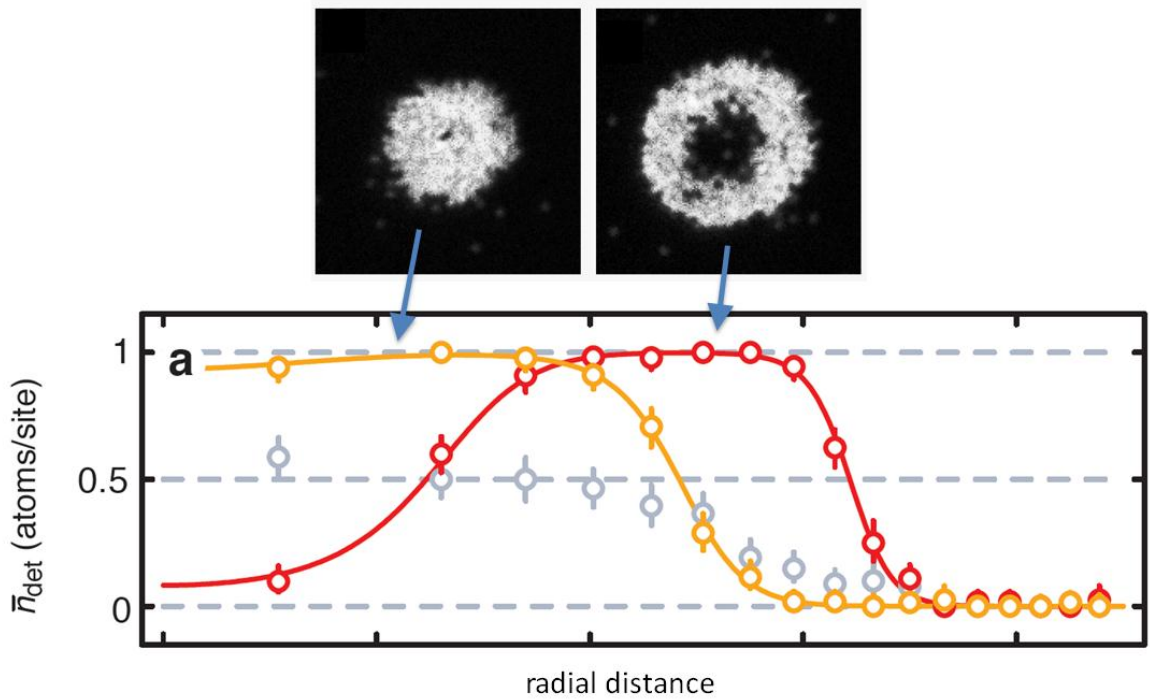

Figure 16. In-situ thermometry via density profiles. In-situ images of a MI with unit filling (left) and two atoms per site (right) are used to determine the average site occupancy at different trap radii. Temperature is determined by fitting to the zero-tunneling approximation (reproduced from Ref. [67]). Reprinted by permission from Macmillan Publishers Ltd: Nature, copyright 2010.

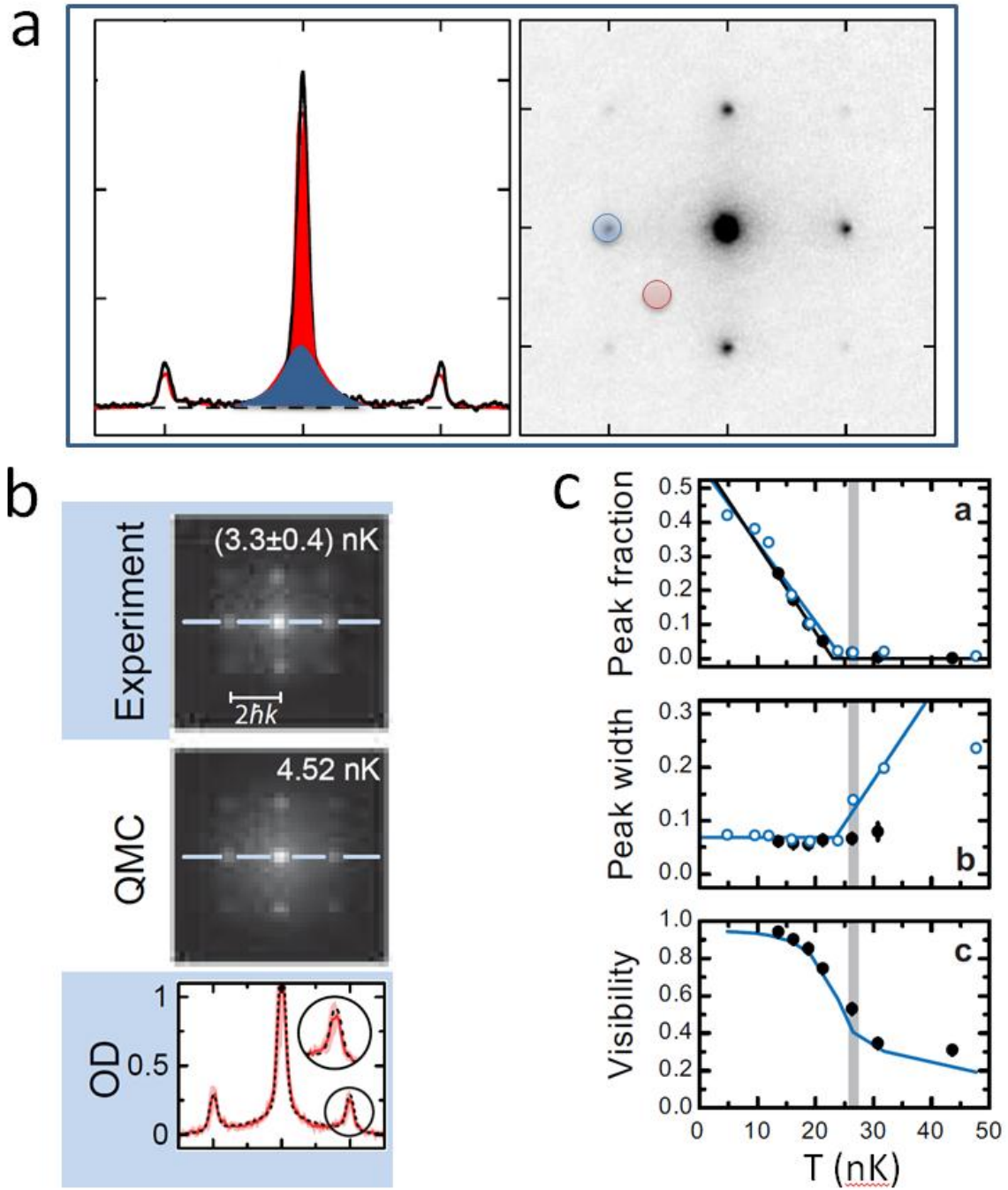

Figure 17. TOF thermometry. Panel (a) shows a typical TOF momentum distribution for a bosonic system (reproduced with permission from Ref. [50], copyright 2008 American Physical Society). A number of properties of such images can be used for thermometry, such as condensate fraction, defined as the fraction of atoms in the narrow (red) component, shown in the cross-section on the left. Visibility, defined as the contrast between the blue and red areas shown in the red, can also be used to determine temperature. Two other methods are shown in (b) and (c) (reproduced from Ref. [134]; reprinted by permission from Macmillan Publishers Ltd: Nature Physics, copyright 2010). In (b), a direct comparison is made to QMC calculations. In (c), heuristic measurements are used to find the critical temperature and/or absolute temperature through comparison to theory.

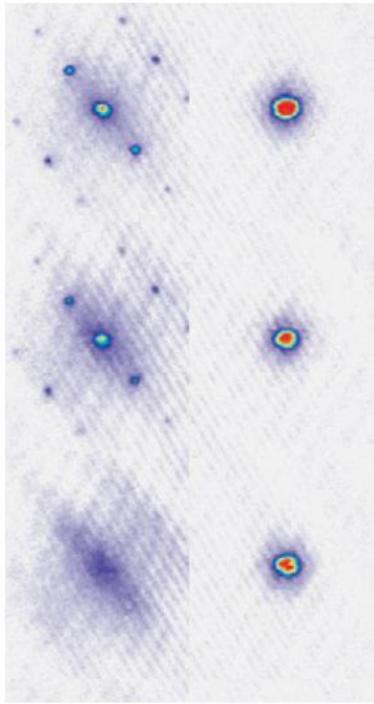 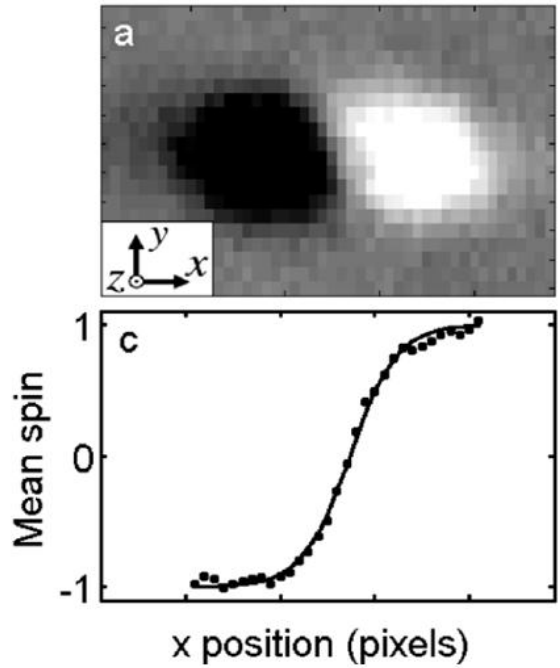

**Figure 18. Extrinsic thermometry.** The panel on the left shows TOF images from a two-component system confined in a spin-dependent lattice. The lattice potential depth is increased from top to bottom. The images for the atoms in the $|1,-1\rangle$ state show this lattice-bound component crossing into the MI regime; the lack of "diffraction peaks" for the $|2,0\rangle$ state show that this co-trapped gas is only parabolically trapped. The weakly interacting gas comprised of atoms in the $|2,0\rangle$ state could be used as an extrinsic thermometer (reproduced from Ref. [47]). An in-situ image for a two-component gas with a magnetic field gradient present is shown in (b). The spin degree of freedom acts as an extrinsic thermometer in thermal contact with the motion, which is the system of interest. The temperature can be determined from the width of the mixing region (reproduced from Ref. [107], copyright 2009 by the American Physical Society).

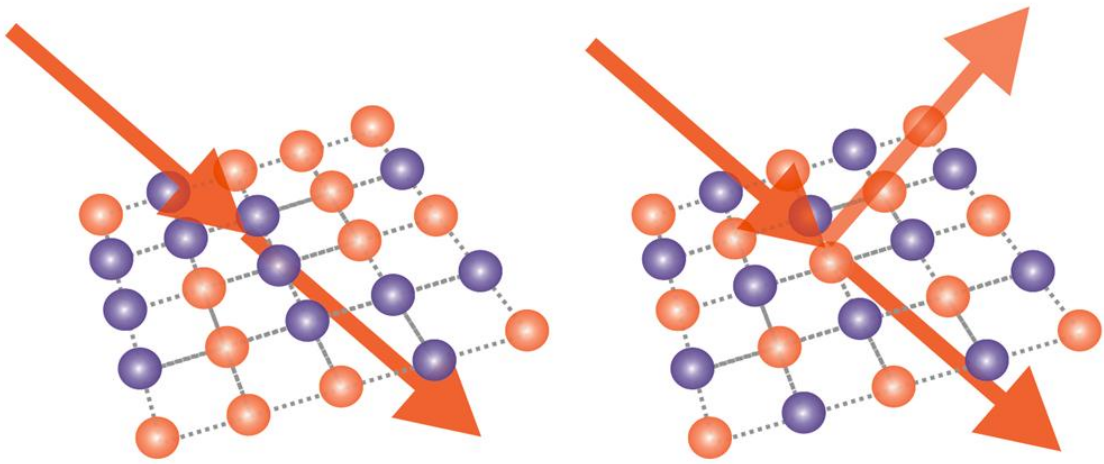

Figure 19. Light-scattering thermometry. Light is scattered from atoms trapped in a lattice in two spin states, indicated by red and blue. The frequency of the light is fixed so that it primarily interacts with atoms in the blue state. In the spin-disordered phase there is no Bragg scattering (left), but in the AFM state the light is deflected. The degree of spin ordering and diffraction is a direct measure of temperature.

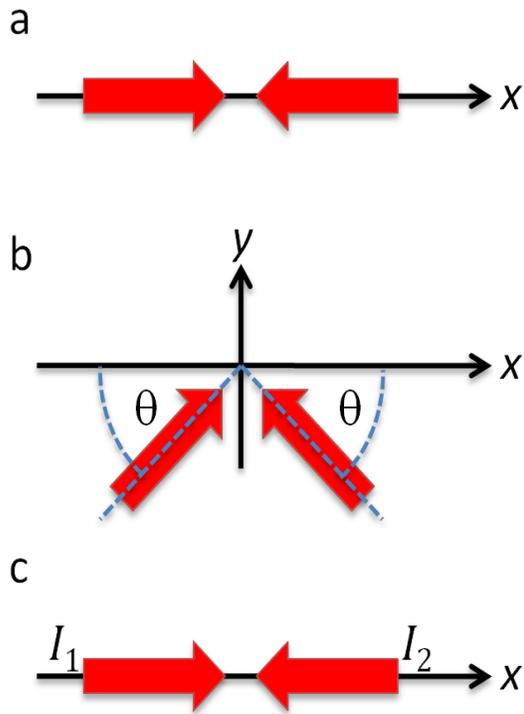

Figure 20. Laser geometries for light-induced heating. In (a) the lattice is formed from counter-propagating beams, in (b) using beams intersecting at an angle, and in (c) using beams with two different intensities.

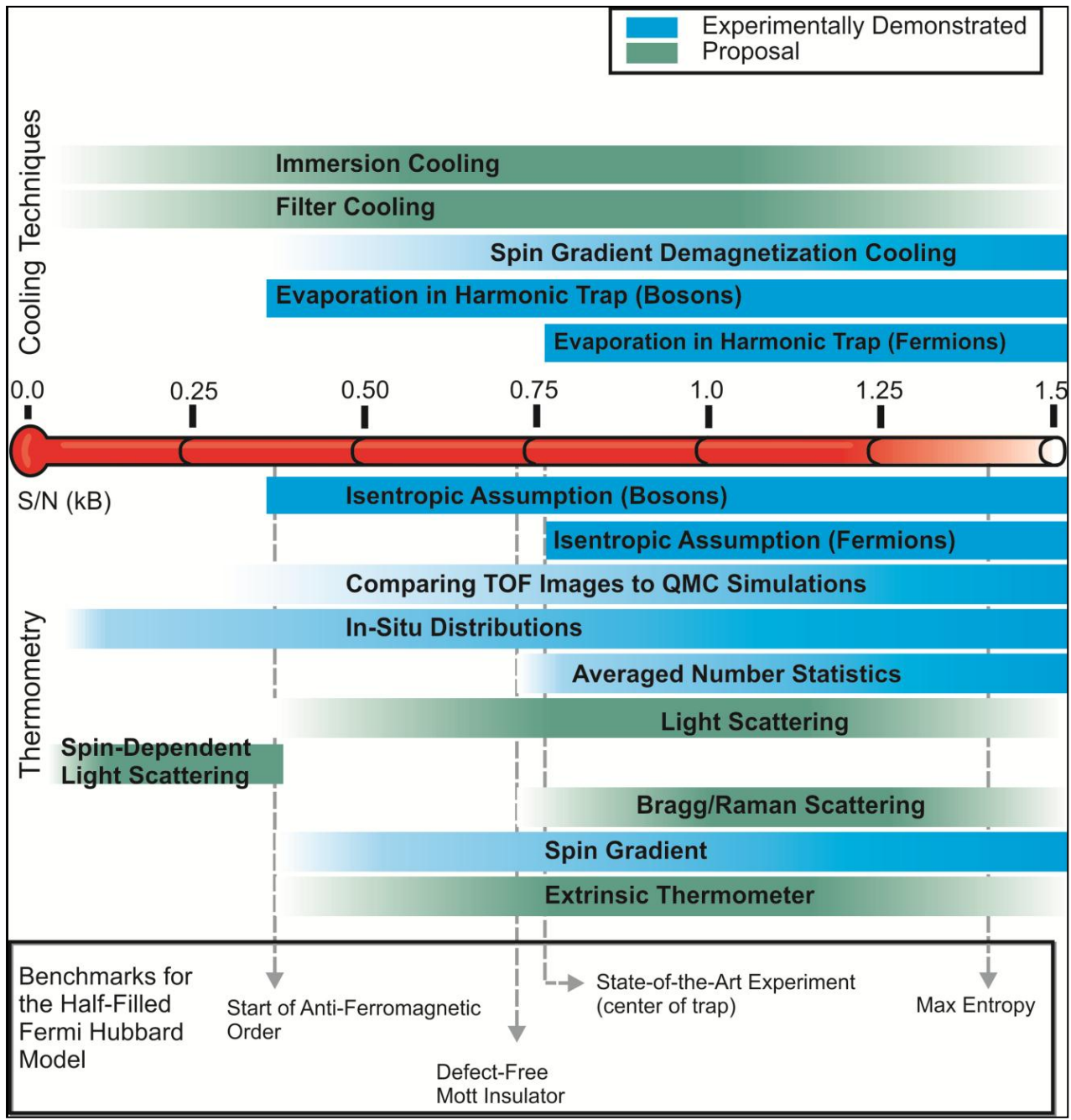

Figure 21: Summary of cooling (top) and thermometry (bottom) methods. Cooling methods are indicated by the range that they are predicted (or demonstrated) to cool to in entropy per particle. Thermometry methods are indicated by their range of validity. Entropies are indicated on the central thermometer, which shows the current limit reached in experiments (red).